\documentclass[a4paper,11pt]{article}
\usepackage{jheppub} 
\usepackage{lineno}
\usepackage{xcolor}
\usepackage{subcaption}
\usepackage{tikz}
\usepackage{orcidlink}

\newcommand{\ket}[1]{\left|#1\right\rangle}
\newcommand{\bra}[1]{\left\langle#1\right|}

\title{\boldmath Toward Hamiltonian simulations of Maxwell-Chern-Simons theory: constant modes and gauge field truncation}

\author[a]{Andrea~Bulgarelli~\orcidlink{0009-0002-2917-6125},}
\affiliation[a]{Transdisciplinary Research Area ``Building Blocks of Matter and Fundamental Interactions'' (TRA Matter) and Helmholtz Institute for Radiation and Nuclear Physics (HISKP), University of Bonn, Nussallee 14-16, 53115 Bonn, Germany}
\emailAdd{abulgare@uni-bonn.de}

\author[b]{Maria~Cristina~Diamantini~\orcidlink{0000-0002-2164-6711},}
\affiliation[b]{NiPS Laboratory, INFN and Dipartimento di Fisica e Geologia, University of Perugia, via A. Pascoli, I-06100 Perugia, Italy}
\emailAdd{cristina.diamantini@pg.infn.it}

\author[a]{Nico~Dichter~\orcidlink{0009-0003-0676-8301},}
\emailAdd{nico@dichter.eu}

\author[a]{Lena~Funcke~\orcidlink{0000-0001-5022-9506},}
\emailAdd{lfuncke@uni-bonn.de}

\author[c,d]{Tobias~Hartung~\orcidlink{0000-0001-6133-5232},}
\affiliation[c]{Northeastern University - London,  Devon House, St Katharine Docks, London, E1W 1LP, United Kingdom}
\affiliation[d]{Khoury College of Computer Sciences, Northeastern University, 440 Huntington Avenue, 202 West Village H Boston, MA 02115, USA}
\emailAdd{tobias.hartung@nulondon.ac.uk}

\author[e,f]{Karl~Jansen,} 
\affiliation[e]{Computation-Based Science and Technology Research Center, The Cyprus Institute,  Nicosia, Cyprus}
\affiliation[f]{Deutsches Elektronen-Synchrotron DESY, Zeuthen, Germany}
\emailAdd{karl.jansen@desy.de}

\author[g,h,i,j]{Enrique~Rico~Ortega~\orcidlink{0000-0003-4414-6821},}
\affiliation[g]{EHU Quantum Center and Department of Physical Chemistry, University of the Basque Country UPV/EHU, P.O. Box 644, 48080 Bilbao, Spain}
\affiliation[h]{DIPC - Donostia International Physics Center, Paseo Manuel de Lardizabal 4, 20018 San Sebastián, Spain}
\affiliation[i]{IKERBASQUE, Basque Foundation for Science, Plaza Euskadi 5, 48009 Bilbao, Spain}
\affiliation[j]{European Organization for Nuclear Research (CERN),  Theoretical Physics Department, CH-1211 Geneva, Switzerland}
\emailAdd{enrique.rico.ortega@cern.ch}

\author[a]{Simran~Singh~\orcidlink{0000-0002-9333-3925},}
\emailAdd{ssingh@uni-bonn.de}

\author[k]{Lorenzo~Spera~\orcidlink{0009-0003-9016-9730}.}
\affiliation[k]{INFN and Dipartimento di Fisica e Geologia, University of Perugia, via A. Pascoli, I-06100 Perugia, Italy}
\emailAdd{lorenzo.spera@studenti.unipg.it}

\abstract{
Maxwell-Chern-Simons (MCS) theory in $2+1$ dimensions provides a paradigmatic example of a topological gauge theory with both dynamical and topological
degrees of freedom. Its Euclidean formulation suffers from a sign problem, making Hamiltonian numerical approaches particularly attractive.
As a first step toward the non-perturbative Hamiltonian study of MCS theory, we investigate the constant-mode sector on a spatial torus. Being analytically solvable in the continuum, it provides an ideal benchmark for understanding how the topological properties of the theory are encoded in a finite-dimensional lattice Hilbert space.
We construct a finite-dimensional discretization of the torus of flat
connections and show that the resulting lattice problem maps onto a generalized Harper-Hofstadter model with twisted boundary conditions.
We identify the commensurability conditions under which the finite lattice exactly reproduces the magnetic translation algebra and the topological degeneracy of the continuum theory. A systematic analysis of gauge field truncation and its convergence toward the continuum limit is then presented.
}

\begin{document}
\maketitle
\flushbottom

\section{Introduction\label{sec:intro}}

Topological gauge theories  play a central role in the modern understanding of topological phases of matter. In particular, in (2+1) dimensions, CS theory provides a universal low-energy description of a wide class of systems, including fractional quantum Hall states and more general topologically ordered phases~\cite{Wen:2016ddy}. However, a purely CS theory does not contain local propagating degrees of freedom~\cite{Chern:1974ft}. The inclusion of a Maxwell term, leading to MCS theory, is therefore essential in order to incorporate local dynamics and to obtain a physically meaningful description beyond the strictly topological limit~\cite{Deser:1982vy}. This necessity of adding a Maxwell term to the CS action becomes even more explicit in lattice formulations, where the Maxwell term plays a crucial regularizing role~\cite{Berruto:2000dp, Berruto:2000fb}.

From a computational perspective, the formulation of MCS theory on a Euclidean lattice is inherently affected by a sign problem induced by the CS term~\cite{Witten:2010cx}. Consequently, not only the real-time dynamics but even the determination of ground state properties becomes highly challenging, as conventional Monte Carlo methods are severely limited. This motivates the formulation of the lattice theory within the Hamiltonian framework, thereby circumventing the sign problem, in such a way that the resulting theory is amenable to tensor network methods and quantum simulation approaches~\cite{Banuls:2019bmf, Funcke:2023jbq, DiMeglio:2023nsa, Davoudi:2026ghz}. A non-exhaustive list of works focusing on Hamiltonian simulations of topological lattice field theories includes~\cite{Magnifico:2018wek, Bermudez:2018eyh, Tirrito:2018bui, Magnifico:2019ulp, Kan:2021nyu, Nakayama:2021iyp, Bharadwaj:2025idp, Rosanowski:2025nck, Huang:2025nlo}.

However, constructing a consistent lattice realization of CS theory is highly nontrivial, as it requires preserving gauge invariance, compactness, and topological properties at finite lattice spacing.
A lattice Hamiltonian formulation has been found and analytically studied in~\cite{Jacobson:2023cmr,Jacobson:2024hov,Peng:2025nfa,Xu:2024hyo,Xu:2026ygx} (for earlier work see also~\cite{Sun:2015hla}). These works, however, rely on the modified Villain formulation of the theory~\cite{Villain:1974ir, Gorantla:2021svj} in order to faithfully represent the compactness of the $U(1)$ gauge group and its topological sectors on the lattice. 

While this approach has led to important progress, it also has a significant consequence: in many implementations, the Villain formulation leads to a Gaussian representation of the theory, which is analytically tractable but may obscure genuinely non-perturbative features. It is therefore  important to find a genuinely non-perturbative framework, where the compact nature of the gauge field is preserved without reducing the dynamics to a Gaussian theory. This is particularly important for tensor-network and quantum-computing applications, where one aims to probe regimes that are not accessible through perturbative expansions.

In this work, we adopt a Hamiltonian perspective from the outset, with the goal of identifying sectors of the theory that can be used as controlled benchmarks for such simulations.
The constant (flat-modes) sector of compact MCS theory on a spatial torus provides exactly a controlled and analytically tractable benchmark for such approaches. 

On a torus,  the gauge field decomposes into flat (harmonic) and non-flat components~\cite{Gukov:2004id}, leading to the factorization of, respectively, the Hamiltonian and the Hilbert space
\begin{equation}
H = H_{\mathrm{nonflat}} + H_{\mathrm{flat}}, \qquad
\mathcal{H} = \mathcal{H}_{\mathrm{nonflat}} \otimes \mathcal{H}_{\mathrm{flat}}.
\end{equation}
The flat sector  corresponds to configurations with vanishing field strength,
\begin{equation}
F = 0,
\end{equation}
and is therefore insensitive to local field-strength fluctuations and to the details of the lattice discretization.
The harmonic modes remain continuous compact variables, identified by large gauge transformations. They satisfy Gauss' law identically and are not removed by local constraints. Their quantization is governed by large gauge transformations and by the global algebra of loop operators.
As a consequence, the topological degeneracy of the theory is entirely carried by the flat sector. The corresponding zero-mode Hamiltonian has the universal form of a particle moving on a compact torus in a uniform effective magnetic field with total flux $2\pi k$ where $k$ is the CS level~\cite{Polychronakos:1989cd,Iengo:1991bgu,Peng:2025nfa}. This leads to a $k$-fold degeneracy of each Landau level, which is a global property and cannot be inferred from the local Hamiltonian density.

In the full quantum theory, odd CS level $k$ requires the specification of a spin structure to define the path integral consistently \cite{Dijkgraaf:1989pz,Belov:2005ze}. This reflects that fermionic CS theories are spin topological quantum field theories, whose relevance to fermionic topological phases of matter, such as the FQHE, has been discussed in \cite{Gaiotto:2015zta,Belov:2005ze}.

In the present work, we fix the boundary conditions from the outset. On a flat torus, this is equivalent to selecting a definite spin structure, since different choices of periodic or anti-periodic boundary conditions along the non-contractible cycles correspond precisely to the different spin sectors.
Within a fixed spin structure, the corresponding spin-dependent factor reduces to a constant phase in the zero-mode sector, where no local topological excitations (such as Dirac-string configurations) are present. This suggests that within the flat sector and at a fixed spin structure, even and odd k can be treated on a similar footing.

The zero-mode sector, therefore, isolates the purely global, topological degrees of freedom of the theory in a form that is both exactly solvable and directly comparable between continuum and lattice formulations.

For concrete numerical simulations, a finite-dimensional Hilbert space is often required in Hamiltonian simulations, in particular those based on tensor networks and digital quantum computing. In the case of the flat modes Hamiltonian, 
this requires a discretization of the zero-mode torus. Importantly, this discretization is conceptually distinct from the lattice regularization of the field theory, as it corresponds to a truncation of global degrees of freedom rather than a discretization of spacetime.

Upon discretization, the problem maps to a finite-dimensional Hofstadter-type system (for a review  see~\cite{Fradkin:2013anc}) with periodic boundary conditions and a necessary gauge patch. The flux per plaquette is:
\begin{equation}
\alpha = \frac{k}{N_x N_y},
\end{equation}
where $N_x, N_y$ are the lattice sizes.

Unlike the standard Hofstadter problem, where the flux per plaquette is fixed and the system size is adjusted accordingly, in the present case the total flux $2\pi k$ is fixed by the CS level, while $\alpha$ depends on the discretization. In the physically relevant regime, corresponding to the continuum limit, one has $k < N_x N_y$, so that $\alpha \ll 1$. In this regime, the effective magnetic unit cell is larger than the lattice spacing, and the implementation of periodic boundary conditions necessarily requires a gauge patch and twisted boundary conditions on the wavefunctions. These features are not optional but are dictated by the nontrivial topology of the underlying U(1) bundle.

A crucial conceptual consequence of this construction is that the spectral structure of the discretized problem exhibits features that are not directly tied to the physical topological degeneracy. In particular, the spectrum typically organizes into clusters of nearly degenerate levels, whose structure depends on arithmetic properties of the discretization (such as the dimensions of the lattice $N_x,$ and $ N_y$, and their relation to $k$).
However, these clusters should not be identified with the topological degeneracy of the continuum theory. The latter is a global property determined solely by the total flux $k$ and is exactly equal to $k$ in the continuum limit. The cluster structure instead reflects the finite-dimensional approximation and disappears in the limit $N_x, N_y \to \infty$ at fixed $k$, where the clusters collapse into exactly degenerate levels.

The paper is organized as follows. In Section~\ref{sec:theory} we establish the mapping between the truncated constant modes and a Hofstadter problem, and we derive the Hamiltonian of our model. Section~\ref{sec:symmetries} is devoted to the study of the relevant symmetries of the truncated constant modes, that are relevant for the analysis of the spectrum. In Section~\ref{sec:numerical_analysis} we use exact diagonalization to numerically study the spectrum of the theory in different regimes, focusing in particular on the continuum limit. Finally, we summarize our results and discuss prospects for simulating the full theory in Section~\ref{sec:conclusions}. Appendix~\ref{app:lattice-hamiltonian-details} provides the detailed transition from the continuum Hamiltonian to the discretized one, while appendix~\ref{app:magnetic-translations-details} provides a detailed derivation of the algebra of magnetic translations.

\section{Derivation of the Hamiltonian\label{sec:theory}}
\subsection{Constant modes Hamiltonian and Hofstadter mapping\label{sec:hofmap}}

We consider a compact $U(1)$ MCS theory defined on a square lattice with periodic boundary conditions and continuous time. The spatial lattice has the topology of a two-dimensional torus. For simplicity, since it will not play any role in what follows, we set the lattice spacing to one and denote by $S$ the total area of the system.

Retaining only the harmonic component of the gauge field, in the temporal gauge $A_0=0$ one has:
\begin{equation}
A_i(t,x) \rightarrow a_i(t), \qquad i=1,2,
\end{equation}
where $a_i(t)$ are spatially constant modes corresponding to the holonomies around the non-contractible cycles of the torus.

The Hamiltonian for the flat modes is:
\begin{equation}
H = \frac{S e^2}{2}\left[\left(p_1 - \frac{k}{4\pi}a_2\right)^2 + \left(p_2 + \frac{k}{4\pi}a_1\right)^2\right],
\label{hfm}
\end{equation}
where $p_i = \pi_i/S$ and $\pi_i$ are the canonical momenta associated with the zero modes. A detailed derivation of this Hamiltonian on the lattice can be found in~\cite{Peng:2025nfa}, while we refer to~\cite{Polychronakos:1989cd} for its derivation in the continuum. 

Equation~\eqref{hfm} can also be obtained starting from the Hamiltonian of compact QED on the lattice, which in the flat sector reduces to
\begin{equation}
H = \frac{S}{2 e^2} (E_1^2 + E_2^2),
\label{hqed}
\end{equation}
with $E_i$ the electric field. The presence of the CS term modifies the relation between the electric field and the canonical momentum as~\cite{Deser:1982vy}
\begin{equation}
\frac{E_i}{e^2} = p_i - \frac{k}{4\pi}\epsilon_{ij} a_j.
\end{equation}
Substituting this relation into the Hamiltonian reproduces Eq.~\eqref{hfm}.
The electric fields satisfy the non-commutative algebra
\begin{equation}
[E_1,E_2]
=
-i\,\frac{e^4 k}{2\pi S}.
\label{Ealgebra}
\end{equation}
The harmonic variables $a_1,a_2$ themselves commute,
\begin{equation}
[a_1,a_2]=0,
\end{equation}
and parametrize the compact zero-mode torus,
\begin{equation}
a_i\sim a_i+2\pi.
\end{equation}
It is convenient to rescale the electric fields according to
\begin{equation}
\tilde E_i=\frac{E_i}{e^2},
\end{equation}
so that
\begin{equation}
[\tilde E_1,\tilde E_2]
=
-i\frac{k}{2\pi S}.
\label{scaledE}
\end{equation}
The structure is completely analogous to the Landau problem: the coordinates commute, while the kinetic momenta do not. The Chern--Simons term therefore induces an effective magnetic field on the compact zero-mode torus.

We now define the combinations
\begin{equation}
a=
\frac{\tilde E_2+i\tilde E_1}{\sqrt{k/\pi S}},
\qquad
a^\dagger=
\frac{\tilde E_2-i\tilde E_1}{\sqrt{k/\pi S}},
\label{ladder}
\end{equation}
which satisfy
\begin{equation}
[a,a^\dagger]=1.
\end{equation}
Using
\begin{equation}
\tilde E_1^2+\tilde E_2^2
=
\frac{k}{\pi S}
\left(
a^\dagger a+\frac12
\right),
\end{equation}
the Hamiltonian becomes
\begin{equation}
H=
\frac{e^2 k}{2\pi}
\left(
a^\dagger a+\frac12
\right).
\label{Landauspectrum}
\end{equation}
The spectrum is therefore
\begin{equation}
E_n=
\frac{e^2 k}{2\pi}
\left(
n+\frac12
\right),
\qquad
n=0,1,2,\dots
\label{Landaulevels}
\end{equation}
namely the standard Landau-level spectrum.

The oscillator algebra alone, however, does not determine the degeneracy of the states. The latter originates from the global topology of the compact zero-mode torus and from the magnetic translation algebra. Defining the operators
\begin{equation}
R_1=\tilde E_2-\frac{k}{2\pi}a_1,
\qquad
R_2=\tilde E_1+\frac{k}{2\pi}a_2,
\end{equation}
one finds
\begin{equation}
[R_i,H]=0, \qquad [R_1, R_2] = -\frac{ik}{2\pi S},
\end{equation}
so that the corresponding magnetic translations commute with the Hamiltonian:
\begin{equation}
U=\exp\left(i\frac{2\pi\sqrt{S}}{k}R_1\right),
\qquad
V=\exp\left(i\frac{2\pi\sqrt{S}}{k}R_2\right).
\end{equation}
They satisfy the Weyl algebra
\begin{equation}
UV=
e^{i2\pi/k}VU.
\label{WeylUV}
\end{equation}
Equation~\eqref{WeylUV} admits a $k$-dimensional irreducible representation. Consequently, each Landau level is $k$-fold degenerate. The full Hilbert space therefore factorizes as
\begin{equation}
\mathcal H
=
\mathcal H_{\rm osc}
\otimes
\mathcal H_{\rm gc}
\qquad
\dim\mathcal H_{\rm gc}=k.
\end{equation}
where $\mathcal H_{\rm osc}$ is the harmonic-oscillator Hilbert space generated by $a^\dagger$, while the finite-dimensional factor $\mathcal H_{\rm gc}$ carries the irreducible representation of the magnetic-translation algebra  Eq.~\eqref{WeylUV}, and describes the guiding-center/topological degeneracy sector.
This reproduces the expected topological degeneracy of compact MCS theory on the torus. 

The appearance of Landau levels reflects the fact that the compact zero-mode sector of MCS theory is equivalent to the motion of a particle on a torus threaded by total magnetic flux
\begin{equation}
\Phi_{\rm tot}=2\pi k.
\end{equation}
Identifying\footnote{Note that both sides of Eq.~\eqref{sym} are gauge fixed. Gauss law does not impose additional constraints on the flat modes $a_i$, while $\tilde{A}_i$ are specified in the symmetric gauge here.
}
\begin{equation}
\widetilde A_1 = -\frac{k}{4\pi} a_2,
\qquad
\widetilde A_2 = \frac{k}{4\pi} a_1,
\label{sym}
\end{equation}
one obtains
\begin{equation}
B_{\text{eff}} = \partial_{a_1}\widetilde A_2 - \partial_{a_2}\widetilde A_1 = \frac{k}{2\pi}.
\end{equation}
The total flux through the $a$-torus is
\begin{equation}
\Phi = \int_{T^2} da_1\, da_2\, B_{\text{eff}} = 2\pi k.
\end{equation}
Consistency under large gauge transformations requires $k \in \mathbb{Z}$. 
The Hamiltonian~\eqref{hfm} describes a particle moving on a torus in the presence of a uniform effective magnetic field in the symmetric gauge \eqref{sym} (see \cite{Al-Hashimi:2008quu} for a detailed treatment). However, the physical properties of a charged particle on a torus threaded by a uniform magnetic flux are not tied to a particular gauge choice. They are determined instead by the total magnetic flux piercing the torus together with the magnetic-translation algebra Eq.~\eqref{WeylUV}.
Different gauge choices therefore provide equivalent continuum descriptions of the same magnetic bundle over the torus: later on we will indeed move to the more convenient Landau gauge.

We now turn to the discretization of the zero-mode torus, which is required for tensor-network and quantum-computing implementations. This discretization is conceptually distinct from the lattice regularization of the field theory, as it corresponds to a finite-dimensional approximation of the zero-mode Hilbert space.
We discretize the torus as:
\begin{equation}
a_1 = \frac{2\pi}{N_x}x = \Delta_x x,
\qquad
a_2 = \frac{2\pi}{N_y}y =\Delta_y y,
\qquad
x=0,\dots,N_x-1,\quad y=0,\dots,N_y-1 ,
\label{discretecoor}
\end{equation}
with $\Delta_x$ and $\Delta_y$ playing the role of lattice spacing.
The discretized Hilbert space has dimension
\begin{equation}
\dim \mathcal{H} = N_xN_y ,
\end{equation}
with basis states $|x,y\rangle$.
The effective flux per plaquette is
\begin{equation}
\alpha = \frac{k}{N_xN_y}.
\end{equation}
At this point, an important distinction with respect to the standard Hofstadter problem should be emphasized. In the latter, the flux per plaquette $\alpha = p/q$ is fixed (with $p$ and $q$ coprime integers), and the lattice size is typically adjusted accordingly~\cite{Fradkin:2013anc}. In the present case, instead, the total flux $2\pi k$ is fixed by the CS level, while the flux per plaquette $\alpha$ depends explicitly on the discretization.
As we will show, this difference leads to a qualitatively distinct spectral structure, in which the topological degeneracy and the lattice-induced orbit structure must be clearly disentangled.

For notational simplicity, in what follows we measure coordinates in units of the lattice spacings $\Delta_x$ and $\Delta_y$, so that $x$ and $y$ denote dimensionless lattice coordinates. In this convention, unit shifts are written as
\begin{equation}
x \to x \pm 1,
\qquad
y \to y \pm 1,
\end{equation}
which correspond to physical shifts
\begin{equation}
a_1 \to a_1 \pm \Delta_x,
\qquad
a_2 \to a_2 \pm \Delta_y.
\end{equation}
Physical units will be restored whenever necessary.

In this representation, the Hamiltonian~\eqref{hfm} can be written in terms of covariant forward and backward lattice derivatives defined by the link variables $U_x(x,y) =  \exp\left(i\,\Delta_x\,\tilde{A}_1\right)$ and $U_y(x,y) =  \exp\left(i\,\Delta_y\,\tilde{A}_2\right)$:
\begin{align}
  D_x^+\psi(x,y) &= \frac{U_x(x)\,\psi( x+\ 1\,y) - \psi(x,y)}{\Delta_x},\label{eq:cov-der+}\\
  D_x^-\psi(x,y) &= \frac{\psi(x,y) - U_x^*(x-1)\,\psi(x-1,y)}{\Delta_x}
\end{align}
where $D^\pm$ are the forward and backward covariant derivatives respectively.
The same holds for $x \leftrightarrow y$. 

For Hamiltonian~\eqref{hfm} we thus have (details are shown in~\ref{app:lattice-hamiltonian-details}):
\begin{align}
    H\psi(i,j)&=\bra{i,j}H\ket{\psi}\nonumber\\
    &=-\frac{e^2S}{2}
    \Bigg[
    \frac{1}{\Delta_x^2}\bigg(U_x(i,j)\psi(i+1,j)+U_x^*(i-1,j)\psi(i-1,j)-2\psi(i,j)\bigg)\nonumber\\
    &\qquad\frac{1}{\Delta_y^2}\bigg(U_y(i,j)\psi(i,j+1)+U_y^*(i,j-1)\psi(i,j-1)-2\psi(i,j)\bigg)
    \Bigg].  \label{hfm1} 
\end{align}

Expanding $\ket{\psi}$ as
\begin{equation}
\ket{\psi} = \sum\limits_{i=0}^{N_x-1}\sum\limits_{j=0}^{N_y-1}\psi(i,j)\ket{i,j},
\label{wann}
\end{equation}

with $\psi(i,j)$ the Wannier state amplitudes~\cite{Fradkin:2013anc}, we find:
\begin{align}
  H&=-\frac{e^2S}{2}
    \sum\limits_{i=0}^{N_x-1}\sum\limits_{j=0}^{N_y-1}\Bigg[\nonumber\\
    &\qquad
    \frac{1}{\Delta_x^2}\bigg(U_x(i,j)\ket{i,j}\bra{i+1,j}+U_x^*(i-1,j)\ket{i,j}\bra{i-1,j}-2\ket{i,j}\bra{i,j}\bigg)\nonumber\\
    &\qquad\frac{1}{\Delta_y^2}\bigg(U_y(i,j)\ket{i,j}\bra{i,j+1}+U_y^*(i,j-1)\ket{i,j}\bra{i,j-1}-2\ket{i,j}\bra{i,j}\bigg)
    \Bigg]
\label{htb}         
\end{align}
The diagonal term $\ket{i,j}\bra{i,j}$ is just a shift in the energies and can be neglected.
Choosing $\Delta_x = \Delta_y = \Delta$ (equal lattice spacing, which implies $N_x = N_y$) one obtains the tight-binding Hamiltonian with hopping parameter\footnote{In the numerical studies we set $e^2 S / 2 = 1$.}
\begin{equation}
  t = \frac{e^2 S}{2\Delta^2}.
\end{equation}
The dicretized variables~\eqref{discretecoor} are compact angular zero-mode coordinates parametrizing the harmonic sector of the gauge field on the torus. Consequently, the quantities conjugate to them are not physical crystal momenta with dimensions of inverse length, but rather dimensionless discrete Fourier phases associated with the finite-dimensional discretization of the compact zero-mode torus.
The corresponding Fourier decomposition is therefore labeled by discrete integers
\begin{equation}
m_x\in \mathbb Z_{N_x},
\qquad
m_y\in \mathbb Z_{N_y},
\end{equation}
with associated dimensionless phases

\begin{equation}
p_x=\frac{2\pi m_x}{N_x},
\qquad
p_y=\frac{2\pi m_y}{N_y}.
\label{momenta}
\end{equation}
These variables take values in the interval
\begin{equation}
0\le p_x,p_y <2\pi,
\end{equation}
and should be interpreted as angular Fourier variables on the compact zero-mode torus rather than physical Bloch momenta.

It is useful to introduce the total flux through the torus
\begin{equation}
  \Phi = 2\pi k = \Phi_0\, k,
\end{equation}
and the flux per plaquette
\begin{equation}
  \Phi_p=\frac{\Phi}{N_x N_y} = \Phi_0\,\frac{k}{N_x N_y} = \Phi_0\alpha  = \Phi_0 \frac{p}{q},
 \end{equation}
with $\Phi_0 = 2\pi,\; p = k, \; q = N_x N_y$.

\subsection{Harper equation and Fourier sectors}
\label{sec:Harper_equation}

We can now switch from the symmetric gauge to the Landau gauge:
\begin{equation}
  \tilde{A}_x = 0, \qquad
  \tilde{A}_y = B\,a_1 = \frac{k}{2\pi}\cdot\frac{2\pi}{N_x}\,x = \frac{k}{N_x}\,x.
\end{equation}
The passage from the symmetric gauge to the Landau gauge does not change the physical content of the zero-mode problem. The Landau problem on a torus is not characterized by a particular gauge choice, but by the total magnetic flux piercing the torus together with the magnetic-translation algebra \cite{Hatsuda:2016mdw,Wakamatsu:2017isl}.
Different gauge choices correspond instead to different local trivializations of the same magnetic $U(1)$ bundle over the torus. Nevertheless, the cocycle condition, the total magnetic flux
\begin{equation}
\Phi_{\rm tot}=2\pi k,
\end{equation}
and the projective magnetic-translation algebra remain unchanged.
The Landau gauge is adopted in the following as it naturally leads to a Harper-type formulation after discretization and it makes the magnetic translation algebra particularly transparent.

In this gauge the link operators become:
\begin{align}
  U_y(x,y) &= \exp\left(i\,\Delta y\,\tilde{A}_y\right) = \exp(2\pi i\,\alpha\,x),\\
  U_x(x,y) &= \exp\left(i\,\Delta x\,\tilde{A}_x\right) = 1.
\end{align}
The Landau gauge is not globally periodic on the torus.
The latter is obtained by identifying lattice sites related by
\begin{equation}
    (x,y)\sim(x+mN_x,y+nN_y),
\qquad m,n\in\mathbb Z.
\end{equation}
A fundamental domain is therefore the rectangular region
\begin{equation}
0\le x\le N_x-1,
\qquad
0\le y\le N_y-1,
\end{equation}
which contains exactly one representative of each equivalence class of the quotient.
If one formally extends the bulk expression
\begin{equation}
U_y(x,y)=\exp(2\pi i\alpha x),
\end{equation}
outside the fundamental domain, one finds
\begin{equation}
U_y(N_x,y)=\exp(2\pi i\alpha N_x)\neq U_y(0,y).
\label{patch}
\end{equation}
This mismatch is precisely the transition function of the magnetic bundle at the boundary. We therefore keep the fundamental domain
\begin{equation}
x=0,\ldots,N_x-1
\end{equation}
and implement the transition function through the boundary link
\begin{align}
  U_x(x,y) &= 1, \quad \forall\, x \neq N_x - 1,\\
  U_x(N_x-1,\,y) &= \exp\!\left(-2\pi i\,\frac{k}{N_y}\,y\right).
  \label{transformed-u-x}
\end{align}
The link crossing the boundary is not identified directly with the bulk link $U_y(0,y)$. Rather, the two are related by the boundary transition function,
\begin{equation}
U_y(N_x,y)
=
g(y)\,
U_y(0,y)\,
g^{-1}(y+1),
\end{equation}
where
\begin{equation}
g(y)=\exp\left(-2\pi i\frac{k}{N_y}y\right).
\label{transfun}
\end{equation}
In the chosen convention, this transition function is absorbed into the twisted boundary link $U_x(N_x-1,y)$. Therefore, the plaquette crossing the boundary is computed entirely within the fundamental domain,using the vertical link in the neighbouring patch, identified with the link at $x=0$,
\begin{equation}
U_y(N_x,y)\ \longrightarrow\ U_y(0,y)=1 .
 \end{equation}
 The flux per plaquette is:
\begin{equation}
  \exp\left(i\Phi_P\right) = U_x(x,y)\,U_y(x+\Delta x,\,y)\,U_x^{-1}(x,\,y+\Delta y)\,U_y^{-1}(x,y).
  \label{flux-landau-gauge}
\end{equation}
In the bulk ($x \neq N_x - 1$) we, thus, have:
\begin{equation}
  \Phi_P = 2\pi \alpha
\end{equation}
At the border $x = N_x - 1$, using \eqref{patch}, we also have:
\begin{equation}
\begin{aligned}
\exp(i\Phi_P(N_x-1,y))
&=
U_x(N_x-1,y)\,
U_y(0,y)\,
U_x^{-1}(N_x-1,y+1)\,
U_y^{-1}(N_x-1,y)
\\
&=
g(y)\,
U_y(0,y)\,
g^{-1}(y+1)\,
U_y^{-1}(N_x-1,y)
\\
&=
\exp\left(-2\pi i\frac{k}{N_y}y\right)
\exp\left(2\pi i\frac{k}{N_y}(y+1)\right)
\exp\left[-2\pi i\alpha(N_x-1)\right]
\\
&=
\exp(2\pi i\alpha),
\end{aligned}
\end{equation}
where we used $\alpha N_x=k/N_y$. Hence the plaquette flux is uniform also at the boundary:
\begin{equation}
\Phi_P(N_x-1,y)=2\pi\alpha .
\end{equation}
To respect gauge invariance under shifts the physical states
must also be transformed accordingly.
Consider the physical state $\ket{x,y}$ at the boundary:
\begin{align}
  \ket{x+N_x,\,y}
    &= \exp(-2\pi i\,\alpha\,N_x\,y)\,\ket{x,y}
     = \exp\!\left(-2\pi i\,\frac{k}{N_y}\,y\right)\ket{x,y},
\label{twist}     
\end{align}
which represents the twisted boundary conditions in $x$: after a complete loop along the $x$ direction the physical state does not remain the same but picks up a phase\footnote{Notice that the explicit form of the twisted boundary conditions depends on the chosen gauge and on the corresponding patching functions. In the present gauge, the twist can be concentrated along one lattice direction, while in the symmetric gauge it is distributed symmetrically between the two directions.}.
Along the $y$ direction, instead, we have:
\begin{equation}
  \ket{x,\,y+N_y} = \ket{x,y}.
\label{notwist}
\end{equation}
Consistency of the boundary conditions \eqref{twist} and \eqref{notwist} applied in different order requires:
\begin{equation}
  \exp(-2\pi i\,\alpha\,N_x N_y) = 1 \;\Rightarrow\; \alpha\,N_x N_y = k \in \mathbb{Z}.
\label{quantiz}
\end{equation}
In the Landau gauge and away from the boundaries, the Hamiltonian density~\eqref{htb} becomes:
\begin{align}
  h(x,y) = -t\Bigl[\ket{x,y}\bra{x+ 1}+ \ket{x,y}\bra{x- 1,y} + e^{2i\pi\alpha x}\ket{x,y}\bra{x,y+ 1}  +e^{-2i\pi \alpha x}\ket{x,y}\bra{x,y- 1}\Bigr],
\label{hlg}  
\end{align}
where we assumed $\Delta_x = \Delta_y = \Delta$; however, for clarity we will write $N_x$ and $N_y$ in what follows.

The eigenvalue equation $H\ket{\psi} = E\ket{\psi}$ yields
\begin{equation}
  E\,\psi(x,y) = -t\Bigl[\psi(x+ 1,\,y) + \psi(x- 1,\,y)
    + e^{2\pi i\alpha x}\psi(x,\,y+ 1)
    + e^{-2\pi i\alpha x}\psi(x,\,y- 1)\Bigr] ,
\label{eigeq}    
\end{equation}
where we used \eqref{wann}.
Since the physical state is periodic in the $y$-direction we can write it as a Fourier decomposition of the form
\begin{equation}
  \psi(x,y) = \frac{1}{\sqrt{N_y}}\sum_{n=0}^{N_y-1} e^{ip_ny}\,g_n(x),
  \qquad  p_n = \frac{2\pi n}{N_y} ,
\label{fexp}  
\end{equation}
and $\displaystyle e^{ip_n}$ are the $N_y$-th root of unitiy. Plugging this expansion in the previous \eqref{eigeq}
we obtain the Harper equation:
\begin{equation}
E\,g_n(x) = -t\Bigl[g_n(x+1) + g_n(x-1 )
+ 2\cos(2\pi\alpha x + p_n)\,g_n(x)\Bigr].
\label{eq:Harper_equation}
\end{equation}
The previous expressions hold in the bulk. To properly treat the twisted boundary, we can use the same Fourier expansion in the $x$ direction
\begin{align}
  &\psi(x+N_x,\,y) = \exp\!\left(-2\pi i\,\frac{k}{N_y}\,y\right)\psi(x,y). \\
  &g_n(x+N_x) = g_{n+k}(x),
  \label{eq:twisted_boundary_conditions}
\end{align}
where the index $n+k$ is understood modulo $N_y$. The full structure of the Hamiltonian then becomes, in components
\begin{align}
    H_{(x,n)\,(x',n')} = 
    \begin{cases}
    -t[\delta_{x',x+1} + \delta_{x',x-1} + 2\cos(2\pi\alpha x + p_n)\delta_{x,x'}]\delta_{n,n'} & \mathrm{if } \;\; 0 < x < N_x-1, \\
    -t[\delta_{x',0}\;\delta_{n',(n+k)\;\mathrm{mod}\; N_y} + \delta_{x',x-1}\delta_{n,n'}] & \mathrm{if} \;\; x = N_x-1, \\
    -t[\delta_{x',x+1}\delta_{n,n'} + \delta_{x',N-1}\;\delta_{n',(n-k)\;\mathrm{mod}\; N_y}] & \mathrm{if} \;\; x = 0,
    \end{cases}
    \label{eq:Harper_Hamiltonian}
\end{align}
where the first row is the bulk term, the other two represent twisted boundaries. In particular $\sum_{(x',n')} H_{(x,n)\,(x',n')}\, g_{n'}(x') = E\,g_{n}(x)$ is the Harper equation. We will refer to~\eqref{eq:Harper_Hamiltonian} as the Harper Hamiltonian.

The condition~\eqref{eq:twisted_boundary_conditions} induces the permutation
\begin{equation}
n \mapsto n+k \pmod{N_y} ,
\end{equation}
of Fourier sectors: after a whole period we do \emph{not} go back to the same sector. One goes back to the original sector after $m$ steps, $m$ chosen such that
\begin{equation}
  m \cdot k \equiv 0 \pmod{N_y}, \qquad \Longrightarrow \qquad m = \frac{N_y}{\gcd(k,N_y)}.
\end{equation}
This permutation partitions the set of Fourier modes into disjoint orbits. The orbit of a given $n_0$ is

\begin{equation}
n_0,\ n_0+k,\ n_0+2k,\ \dots \pmod{N_y}.
\end{equation}
The orbit closes after the smallest $\ell_{\mathrm{robit}} > 0$ such that
\begin{align}
    k\ell_{\mathrm{orbit}} \equiv 0,
\end{align}
namely
\begin{equation}
 \ell_{\mathrm{orb}} = \frac{N_y}{\gcd(k, N_y)},
 \label{lorb}
\end{equation}
is the length of the orbit. The number of distinct orbits is
\begin{equation}
 N_{\mathrm{orb}} = \gcd(k, N_y) = \frac{N_y}{\ell_{\mathrm{orb}}}.
 \label{norb}
\end{equation}
This structure is purely kinematical: it follows directly from the boundary conditions. It classifies how Fourier sectors mix under a winding in the $x$-direction, but it does not yet say anything about spectral degeneracy.

\section{Symmetries of the model}
\label{sec:symmetries}
\subsection{Magnetic Translations \label{sec:Magnetic_Translations}}
We define the \textit{magnetic translation operators} to be
\begin{align}
  T_x\ket{x,y} &= \exp(-2\pi i\,\alpha\,y)\ket{x+ 1,\,y},\\
  T_y\ket{x,y} &= \ket{x,\,y+ 1} ,
  \label{mtra}
\end{align}
and they satisfiy the magnetic algebra:
\begin{equation}
  T_x T_y = \exp(2\pi i\,\alpha)\,T_y T_x .
  \label{eq:mt-algebra}
\end{equation}
The complete algebra can be found in Appendix~\ref{app:magnetic-translations-details}.
While in the bulk the operators \eqref{mtra} commute with the Hamiltonian, this is not true at the border. This reflects a global obstruction in defining magnetic translation in our setting.
The global obstruction to defining periodic magnetic translations is not a peculiarity of the Landau gauge. It follows directly from the projective magnetic-translation algebra together with the compactness of the torus. The obstruction is therefore gauge independent and reflects the nontrivial topology of the magnetic $U(1)$ bundle carrying total flux $2\pi k$.

On a torus, consistency requires that going around a non-contractible cycle acts trivially on the
Hilbert space:
\begin{equation}\label{eq:global}
(T_x)^{N_x}=1,\qquad (T_y)^{N_y}=1.
\end{equation}
If $k\neq 0$, the conditions \eqref{eq:global} cannot be simultaneously satisfied by operators obeying
\eqref{eq:mt-algebra}.
Using \eqref{eq:mt-algebra} repeatedly,
\begin{equation}
(T_x)^{N_x}T_y = e^{i2\pi\alpha N_x}\,T_y (T_x)^{N_x}.
\end{equation}
If $(T_x)^{N_x}=1$, then the left-hand side equals $T_y$, hence
\begin{equation}
T_y = e^{i2\pi\alpha N_x}\,T_y.
\end{equation}
Therefore $e^{i2\pi\alpha N_x}=1$, i.e.\ $\alpha N_x\in\mathbb{Z}$. Since
$\alpha=\frac{k}{N_xN_y}$, this requires $\frac{k}{N_y}\in\mathbb{Z}$, i.e.\ $N_y\mid k$.
Exchanging $x\leftrightarrow y$ yields also $N_x\mid k$. For generic sizes this fails, hence
\eqref{eq:global} is obstructed. Even in commensurate cases, the full algebra cannot be realized
globally for the fundamental generators $T_x,T_y$; only suitable powers close.
Although the fundamental generators cannot satisfy \eqref{eq:global}, suitable powers do.
To see this, we define $g_y=\gcd(k,N_y)$ and $m_y=N_y/g_y$. Then
\begin{equation}\label{eq:reduced}
\bigl((T_x)^{N_x}\bigr)^{m_y}=1.
\end{equation}
Analogously, with $g_x=\gcd(k,N_x)$ and $m_x=N_x/g_x$, one has
\begin{equation}
\bigl((T_y)^{N_y}\bigr)^{m_x}=1.
\end{equation}
From the commutation relation \eqref{eq:mt-algebra} one computes
\begin{equation}
(T_x)^{N_x}T_y = e^{i2\pi \frac{k}{N_y}}\,T_y (T_x)^{N_x}.
\end{equation}
Iterating $m_y$ times gives
\begin{equation}
\bigl((T_x)^{N_x}\bigr)^{m_y} T_y = e^{i2\pi k\frac{m_y}{N_y}}\,T_y \bigl((T_x)^{N_x}\bigr)^{m_y}
= e^{i2\pi \frac{k}{g_y}}\,T_y \bigl((T_x)^{N_x}\bigr)^{m_y} = T_y \bigl((T_x)^{N_x}\bigr)^{m_y},
\end{equation}
since $k/g_y\in\mathbb{Z}$. The same holds with $T_x$ in place of $T_y$, hence the operator
$\bigl((T_x)^{N_x}\bigr)^{m_y}$ commutes with the algebra and can be taken to be the identity in
an irreducible representation.

Only an \emph{arithmetic subgroup} of magnetic translations survives globally. This is the lattice
counterpart of the fact that only a discrete subgroup of translations is compatible with twisted
boundary conditions and reflects the orbits structure.
The obstruction is topological: it reflects the non-triviality of the $U(1)$ bundle over the torus
with first Chern number $k$. Wavefunctions are sections with \emph{twisted} boundary conditions.

Both $(T_x)^{N_x \cdot N_y} = 1$ and $(T_y)^{N_x \cdot N_y} = 1$.
These conditions are reflected in the fact that the total flux through the configuration of the torus is a multiple of $\Phi_0 = 2\pi$.

\subsection{Chiral symmetry}
\label{sec:chiral_symmetry}
We study additional symmetries of the model, originating from the structure of the Harper Hamiltonian in Eq.~\eqref{eq:Harper_equation}, and analyze its consequences for the
spectrum. Specifically, we show that when both $N_x$ and $N_y$ are even,
the model possesses a chiral symmetry generated by a unitary, Hermitian
operator $\Gamma$ satisfying
\begin{equation}\label{eqn:chi_sym}
    \{\Gamma, H\} = 0\,, \qquad \Gamma^2 = \mathbb{I}\,,
\end{equation}
An immediate consequence is a spectrum symmetric about zero energy: if
$\ket{\psi}$ is an eigenstate with energy $E$, then $\Gamma\ket{\psi}$ is an
eigenstate with energy $-E$. Although an operator satisfying
Eq.~\eqref{eqn:chi_sym} is not unique, exhibiting a single such operator
suffices to establish the symmetry. Based on the bipartite structure of the 2D Hofstadter Hamiltonian (section~\ref{sec:hofmap}) to which we have mapped our problem and independently based on our numerical experiments, which show that, for $N_x=N_y=N$, the spectrum is symmetric only when $N$ is even, we adopt the ansatz
\begin{equation}\label{eqn:compact_Gamma}
    \Gamma = \mathrm{diag}_x((-1)^x) \otimes \, T^{p_y}_{N/2} \,.
\end{equation} 
In component form, this operator becomes,
\begin{align}\label{eqn:comp_Gamma}
    \Gamma_{(x,n),(x',n')} = (-1)^x\,\delta_{x,x'}\, \delta_{n',(n+N)\,\mathrm{mod}\, N}\,,
\end{align}
displaying a staggered sign change along the $x$-direction, and a half-lattice translation in the $y$-direction in momentum space. Importantly, the translation is defined only for even $N$. As we show below, the obstruction to creating such an operator for odd lattices comes from two different effects: $N/2$ is not an integer, so the lattice translation is not well defined. And at the boundary the phase from the staggered transformation in the $x$ direction is $+1$, due to $N_x$ being odd.

In the following we show that the operator defined in Eq.~\eqref{eqn:comp_Gamma}, satisfies Eq.~\eqref{eqn:chi_sym} by computing the following sequence of operations: $\Gamma H \Gamma$ on a state-vector $g_n(x)$. We show that the result of this operation returns $-H g_n(x)$  -- verifying $\Gamma$ as a valid chiral operator.

\begin{figure}[t]
    \centering
    \begin{subfigure}{.45\linewidth}
    \includegraphics[width=1\linewidth]{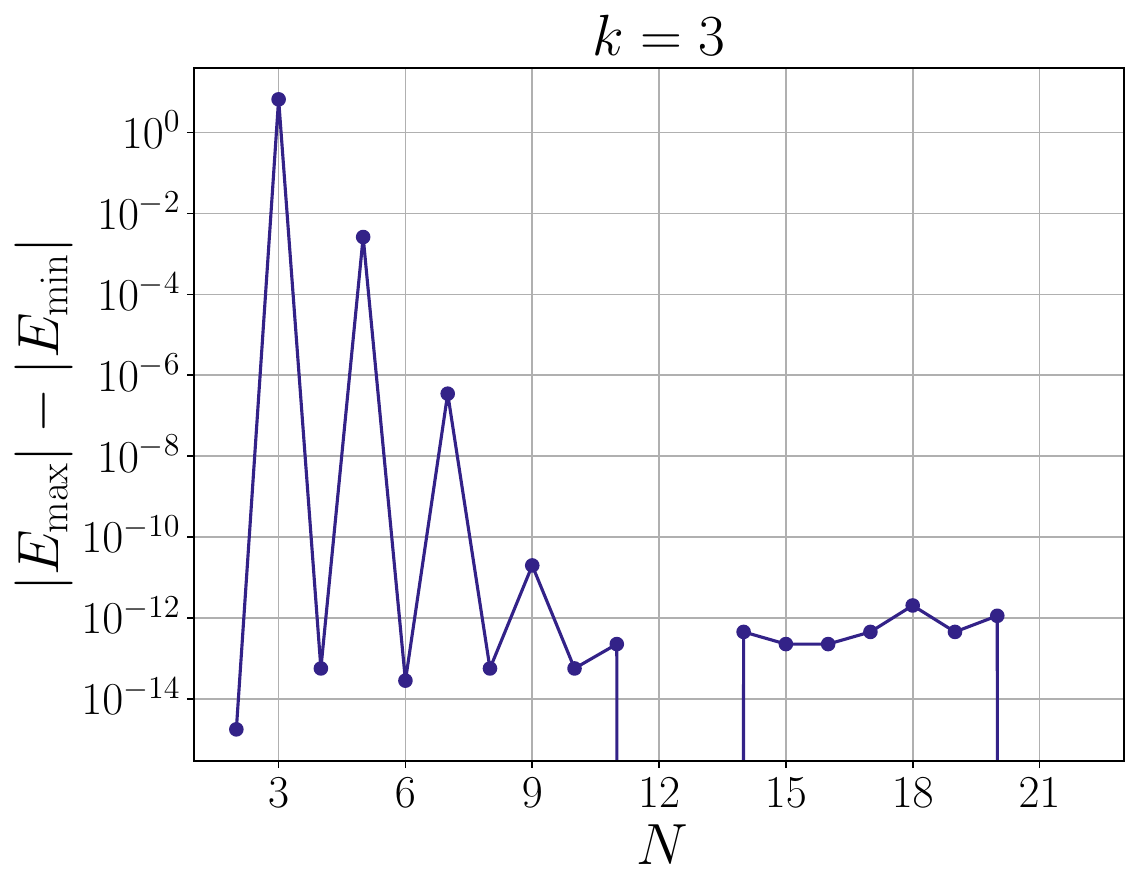}
    \caption{}
    \end{subfigure}
    \begin{subfigure}{.45\linewidth}
    \includegraphics[width=1\linewidth]{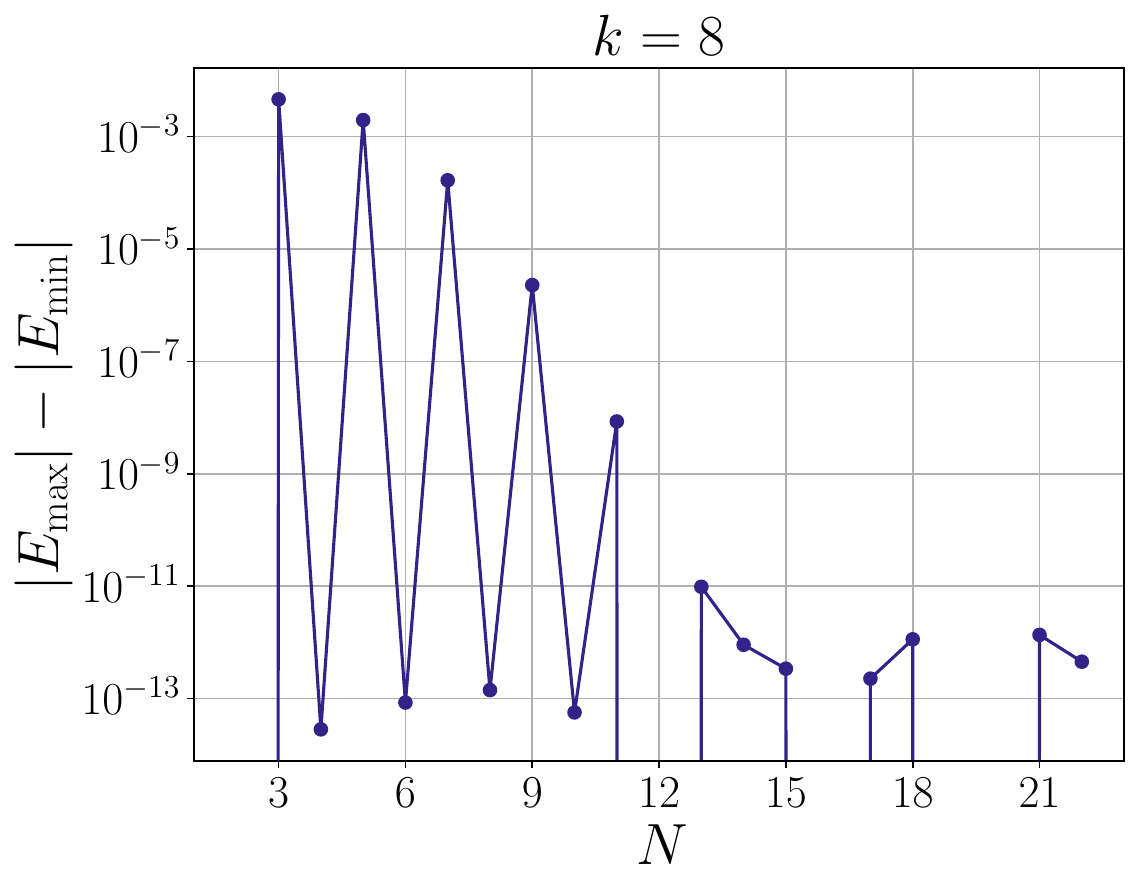}
    \caption{}
    \end{subfigure}
    \caption{Difference between the absolute values of the ground state energy and the highest excited-state energy as a function of system size for (a) $k=3$ and (b) $k=8$. Within numerical precision, the spectrum is symmetric for even $N$.}
    \label{fig:symmetry_spectrum}
\end{figure}

The action of the chiral operator on a state-vector $g_n(x)$ is
\begin{align}
    \sum_{(x',n')}\Gamma_{(x,n)(x',n')}\,g_{n'}(x') = (-)^x \, g_{n+N/2}(x).
\end{align}
Applying the Harper Hamiltonian~\eqref{eq:Harper_Hamiltonian} now leads to
\begin{align*}
\sum_{(x',n')}H_{(x,n)(x',n')}&\left[(-)^{x'} g_{n'+N/2}(x') \right] = \\
    &\begin{cases}
        -t \bigl[ (-)^{x+1}g_{n+N/2}(x+1) + (-)^{x-1}g_{n+N/2}(x-1) \\
        \quad + 2\cos(2\pi\alpha x + p_n)\, g_{n+N/2}(x)\bigr]
            & \text{if } 0 < x < N-1, \\[4pt]
        -t\bigl[(-)^{0}g_{n+N/2+k}(0) + (-)^{N-2}g_{n+N/2}(N-2)\bigr]
            & \text{if } x = N-1, \\[4pt]
        -t\bigl[(-)^1 g_{n+N/2}(1) + (-)^{N-1}g_{n+N/2-k}(N-1)\bigr]
            & \text{if } x = 0.
    \end{cases}
\end{align*}
Finally, further applying $\Gamma^{-1} = \Gamma$ one obtains
\begin{align*}
    \begin{cases}
        -t\bigl[(-)^{x+1}(-)^x g_n(x+1) + (-)^{x-1}(-)^x g_n(x-1) \\
        \quad + 2\cos(2\pi\alpha x + p_n + \pi)\,g_{n}(x)\bigr]
            & \text{if } 0 < x < N-1, \\[4pt]
        -t\bigl[(-)^{N-1}(-)^0\, g_{n+k}(0) + (-)^{N-1}(-)^{N-2}g_n(N-2)\bigr]
            & \text{if } x = N-1, \\[4pt]
        -t\bigl[(-)^0(-)^{1}g_{n}(1) + (-)^{0}(-)^{N-1}g_{n-k}(N-1)\bigr]
            & \text{if } x = 0.
    \end{cases}
\end{align*}
The result of this operation is nothing but $-H g_n(x)$. The equality holds
term by term, for the following reasons. The on-site term changes sign because
the half-shift gives $p_{n+N_y/2} = p_n + \pi$, so that
$\cos(2\pi\alpha x + p_n + \pi) = -\cos(2\pi\alpha x + p_n)$. This requires
$N_y$ to be even, otherwise $N_y/2$ is not an integer and the shifted sector
does not exist. The hopping terms change sign through the staggered factor: in
the bulk one has $(-1)^x(-1)^{x\pm 1} = -1$, while at the boundary the
wrap-around bond connects $x = N_x - 1$ to $x = 0$ and gives
$(-1)^{N_x-1}(-1)^0 = (-1)^{N_x-1}$, which is equal to $-1$ only for $N_x$
even. Finally, two applications of $\Gamma$ shift the transverse index by
$N_y$, which acts trivially since $g_{n+N_y}(x) = g_n(x)$, and at the boundary
$g_{n+N_y\pm k}(x) = g_{n\pm k}(x)$, reproducing the couplings of $H$. Both obstructions to constructing such an operator for odd lattices can thus be seen as a consequence of the fact that an odd square lattice is not bipartite.

In Fig.~\ref{fig:symmetry_spectrum} we numerically probe the chiral symmetry in the spectrum. For different values of $N$, we plot the difference between the absolute values of the ground state energy, $E_{\mathrm{min}}$, and the highest excited-state energy, $E_{\mathrm{max}}$. For even values of $N$ we consistently find, within numerical precision, a symmetric spectrum, while for odd values of $N$ the symmetry appears to be explicitly broken, and only recovered for large values of $N$.

\section{Numerical analysis of the spectra}
\label{sec:numerical_analysis}
\subsection{Continuum limit and admissible discretizations}
\label{sec:continuum_limit}

\begin{figure}
    \centering
    \begin{subfigure}{.48\linewidth}
    \includegraphics[width=1\linewidth]{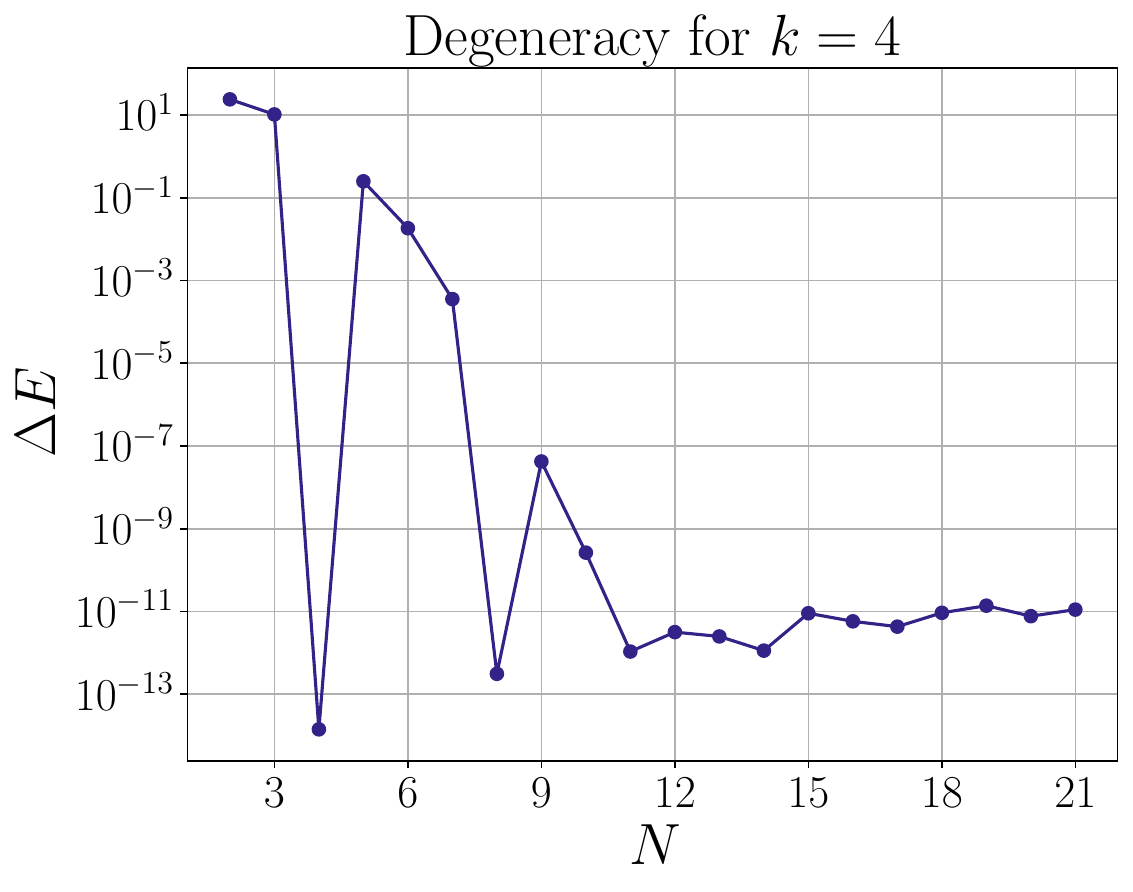}
    \caption{}
    \end{subfigure}
    \begin{subfigure}{.48\linewidth}
    \includegraphics[width=1\linewidth]{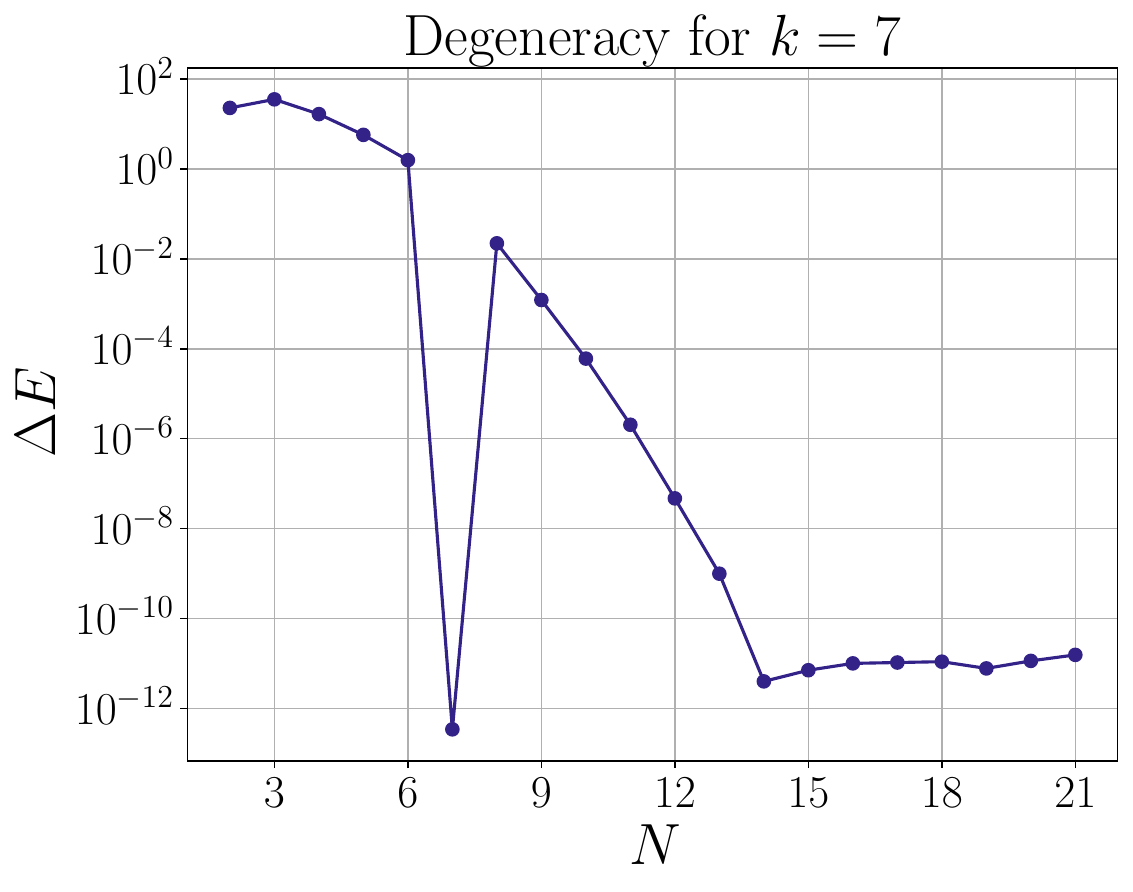}
    \caption{}
    \end{subfigure}
    \caption{The degeneracy of the energy levels is approached exponentially fast as $N$ is increased. In commensurate cases, the degeneracy becomes exact.}
    \label{fig:convergence_degeneracy}
\end{figure}

The continuum limit is obtained by taking
\begin{equation}
N_x, N_y \to \infty, \qquad \alpha = \frac{k}{N_x N_y} \to 0,
\label{contlim}
\end{equation}
at fixed $k$.

Let $H_{N_x,N_y}$ be the lattice Hamiltonian with $\alpha = \frac{k}{N_x N_y}$. As
$N_x, N_y \to \infty$ (at fixed total flux), $H_{N_x,N_y}$ converges to the continuum Landau Hamiltonian $H$ with flux $2 \pi k$. This leads to an emergent $k$-fold quasi-degeneracy: for each Landau level $n$, there exist $k$ eigenvalues $E_{n,1}, \dots, E_{n,k}$ of $H_{N_x,N_y}$ such that
\begin{equation}
\lim_{N_x,N_y\to\infty} |E_{n,a}-E_{n,b}| = 0 \qquad \forall a,b.
\end{equation}
In other words, eigenvalues of $H_{N_x,N_y}$ near a degenerate eigenvalue of $H$ organize into clusters of $k$ levels whose spread vanishes in the limit. This degeneracy is \emph{dynamical}: it is not enforced by the discrete algebra, but emerges in the continuum limit.

For any finite discretization, the spectrum contains only a finite number of states. The continuum Landau spectrum is recovered progressively in the limit $N_x,N_y\to\infty$, where an increasing number of low-energy states organize into quasi-degenerate $k$-plets.

The agreement with the continuum Landau-level structure is most accurate for the lowest states, which probe length scales much larger than the lattice spacing and are therefore less sensitive to discretization effects.
Higher-energy states are increasingly affected by lattice artifacts, since they probe shorter length scales and are more sensitive to the underlying discrete structure.

Numerical evidence for this convergence is presented in Fig.~\ref{fig:convergence_degeneracy}. Specifically, the plots show the difference between the largest and the smallest energy within the first $k$ states, as a function of $N$. The $k$-fold degeneracy is approached exponentially fast, up to values of $N$ where the degeneracy is exact even at finite lattice spacing. A detailed analysis of these special points will be presented in Section~\ref{subsec:case_II}.

\begin{figure}
    \centering
    \begin{subfigure}{.48\linewidth}
    \includegraphics[width=1\linewidth]{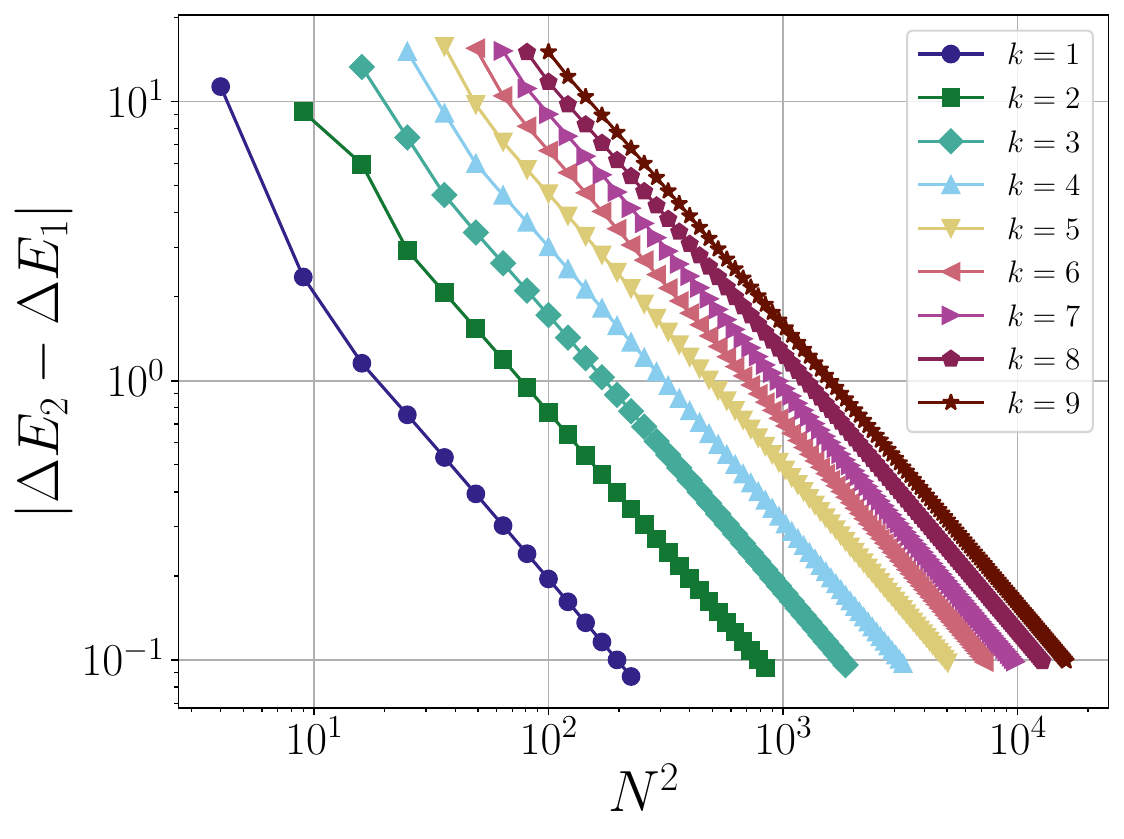}
    \caption{}
    \end{subfigure}
    \begin{subfigure}{.48\linewidth}
    \includegraphics[width=1\linewidth]{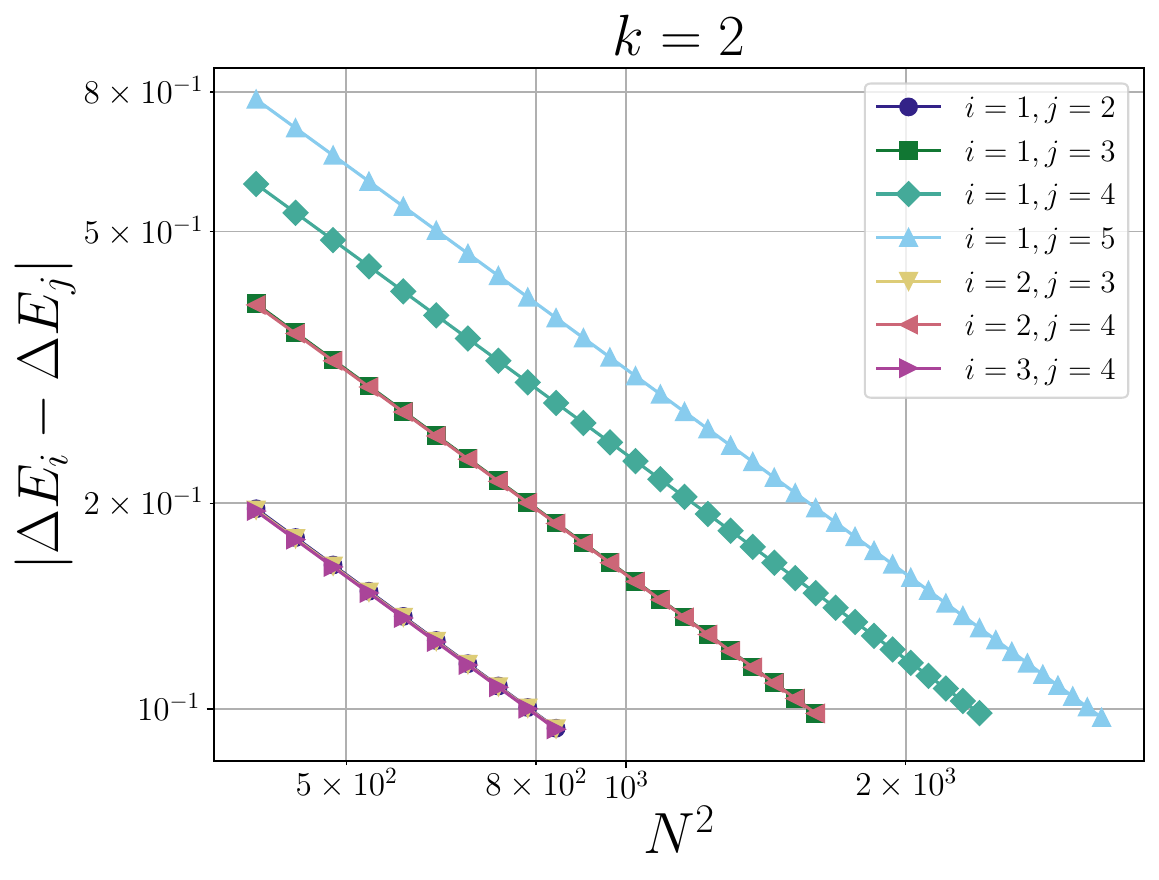}
    \caption{}
    \end{subfigure}
    \caption{Convergence of the level spacing to a harmonic oscillator for different values of the CS level $k$. $\Delta E_{i}$ represents the energy gap between the $(i-1)$-th and the $i$-th quasi-degenerate multiplets.}
    \label{fig:level_spacing_convergence}
\end{figure}

In Fig.~\ref{fig:level_spacing_convergence} we also investigate the level spacing as the continuum limit is approached. In the continuum theory, the Landau levels are equally spaced, since the Hamiltonian, up to degeneracy, reduces to that of a harmonic oscillator. In the left panel, we compare the first and the second level spacing for different values of\footnote{We define $\Delta E_i$ as the energy difference between the highest energy level within the $(i-1)$-th multiplet and the lowest energy in the $i$-th multiplet, with $0$ being the quasi-degenerate ground state multiplet.} $N^2$. The results exhibit a clear algebraic convergence toward the continuum behavior. Our data are well approximated by the functional form $A / (N^2)^{\alpha}$; imposing $\alpha$ as a global fit parameter we obtain the estimate $\alpha=1.039(5)$.

In the right panel of Fig.~\ref{fig:level_spacing_convergence}, we show evidence that also higher energy levels converge to a harmonic oscillator. In particular, we plot the difference between the $i$-th and the $j$-th energy level. Deviations from the harmonic oscillator spectrum seem to depend mostly on $|i-j|$.

Overall, our results are promising in light of simulating the full MCS theory. Away from the continuum limit, the many-body dynamics is expected to be dominated primarily by the low-lying states of the individual lattice d.o.f. (see, for example, Ref.~\cite{Bruckmann:2018usp, Tong:2021rfv}). In the constant-mode sector, we have shown that these states reproduce the corresponding continuum physics with exponential accuracy as the truncation is increased. By contrast, the convergence of the higher-energy spectrum toward the harmonic-oscillator behavior is slower and follows an algebraic scaling. Nevertheless, excited states become increasingly relevant only in the vicinity of the continuum limit. As the system approaches criticality, a larger portion of the local Hilbert space is explored and truncation effects naturally become more pronounced. Whether this ultimately constitutes a limitation depends on the extent to which continuum physics can be reliably extrapolated from simulations performed at finite lattice spacing. While our results are not conclusive for the full theory, the asymptotic scaling of the level spacing is already clearly visible at relatively small values of $N$, suggesting that the approach to the continuum regime can be quantitatively controlled even at moderate truncation levels.

Our findings imply that a faithful discretization of the zero-mode sector requires $\alpha \ll 1$. Configurations outside this regime do not resolve the magnetic structure and should be excluded in numerical implementations. However, this condition alone is not sufficient to guarantee the correct continuum limit. An additional requirement must be satisfied in order for the $k$-fold topological degeneracy to emerge:
\begin{equation}
k < N_x, \qquad k < N_y.
\label{cond}
\end{equation}
This condition reflects the fact that, at fixed total flux, the discretization must be fine enough to resolve the underlying magnetic structure.

A semiclassical argument helps to clarify the underlying physics. For $\alpha \ll 1$, in the Harper Eq.~\eqref{eq:Harper_equation}, the cosine term
\begin{equation}
V_n(x) = 2\cos(2\pi \alpha x + p_n)
\end{equation}
can be viewed as a slowly varying potential. For fixed $n$, its minima occur at
\begin{equation}
2\pi \alpha x_0 + p_n = (2m+1)\pi, \qquad m\in\mathbb{Z},
\label{cospot}
\end{equation}
which gives
\begin{equation}
x_0(n,m) = \frac{(2m+1)\pi - p_n}{2\pi \alpha}.
\label{guiding}
\end{equation}
On the torus, these satisfy
\begin{equation}
x_0(n+k,m) \equiv x_0(n,m) \pmod{N_x}.
\end{equation}

The quantities $x_0(n,m)$ can be interpreted as lattice analogues of the guiding centers in the continuum Landau problem. The Landau-level degeneracy originates from the non-commuting guiding center coordinates, which define a set of zero-energy degrees of freedom. The number of independent guiding centers is equal to the total magnetic flux, yielding a degeneracy $\Phi/2\pi$~\cite{Al-Hashimi:2008quu}.

The minima associated with consecutive Fourier sectors are separated by
\begin{equation}
x_0(n+1,m) - x_0(n,m) = -\frac{N_x}{k}.
\label{deltamin}
\end{equation}
Thus, the $k$ guiding centers are well resolved only when $N_x/k \gg 1$, i.e.\ for $k \ll N_x$. When $k > N_x$, even though $\alpha < 1$, the discretization is too coarse to resolve the individual guiding centers, and the quasi-degenerate $k$-plet is no longer clearly visible.

Configurations with $\alpha = \mathcal{O}(1)$, and in particular $\alpha \in \mathbb{Z}$, do not belong to the continuum scaling regime of the zero-mode sector. They correspond to coarse discretizations in which the topological structure of the theory is not resolved. Therefore, such cases are not relevant for reproducing the continuum MCS physics and should be excluded in numerical implementations. Nevertheless, they exhibit interesting spectral features, which we will briefly discuss.

\subsubsection{Case I: $\gcd(k,N)=1$ with $k < N$}

\begin{figure}
    \centering
    \begin{subfigure}{.48\linewidth}
    \includegraphics[width=1\linewidth]{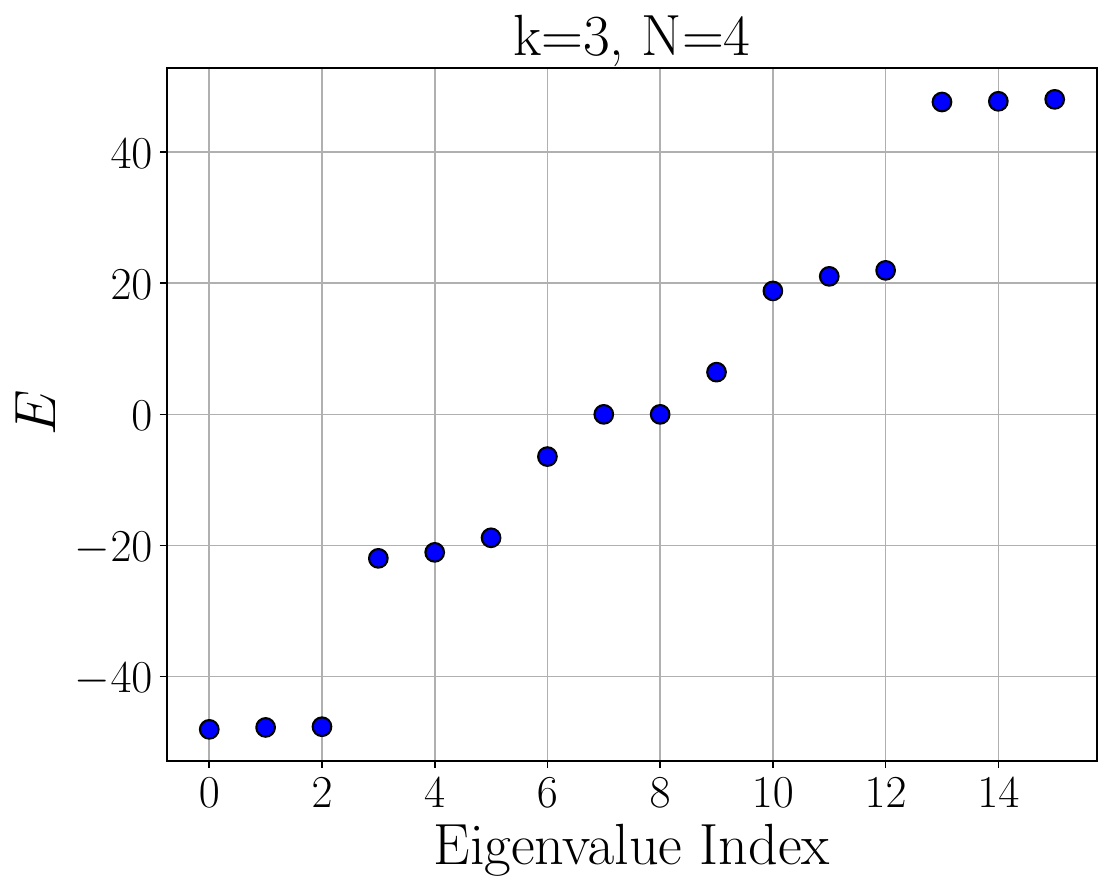}
    \caption{}
    \end{subfigure}
    \begin{subfigure}{.48\linewidth}
    \includegraphics[width=1\linewidth]{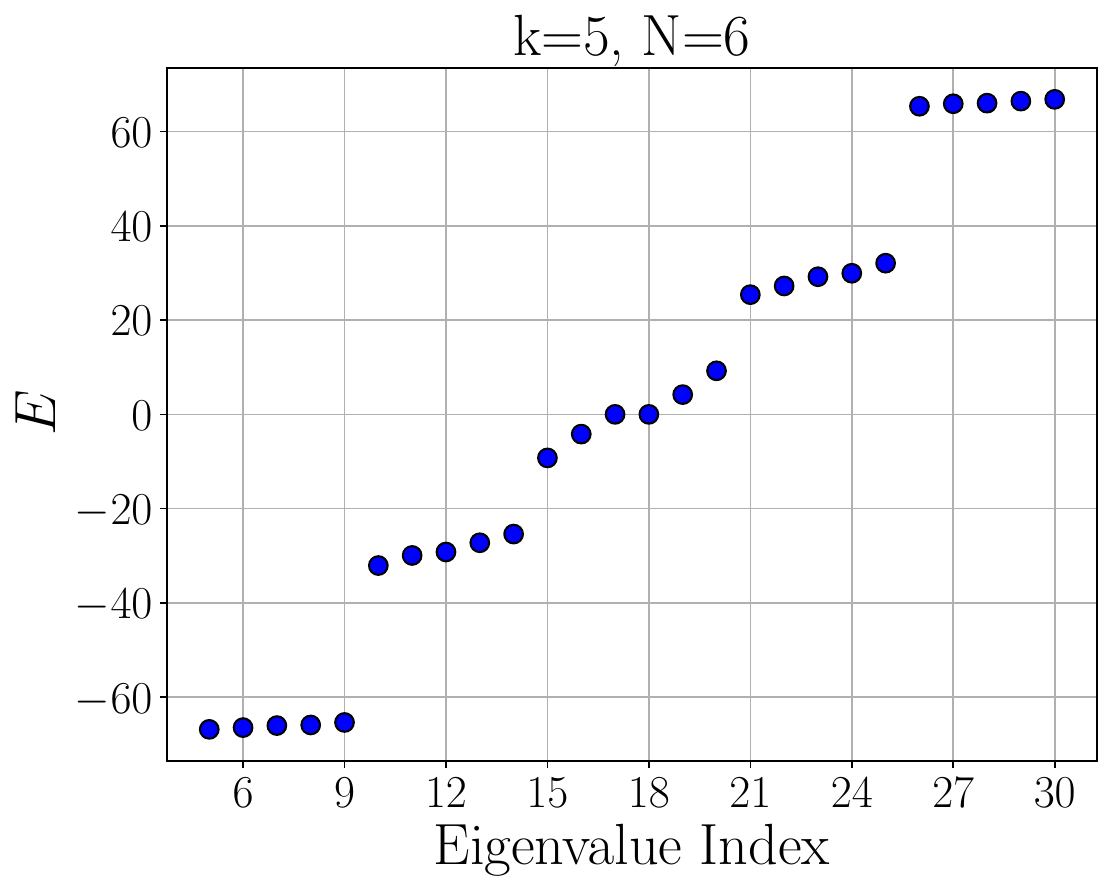}
    \caption{}
    \end{subfigure}
    \caption{Mid spectrum for (a) $k=3$, $N = 4$ and (b) $k=5$, $N = 6$.}
    \label{fig:mid_spectrum_k_odd_N_even}
\end{figure}

\begin{figure}
    \centering
    \begin{subfigure}{.48\linewidth}
    \includegraphics[width=1\linewidth]{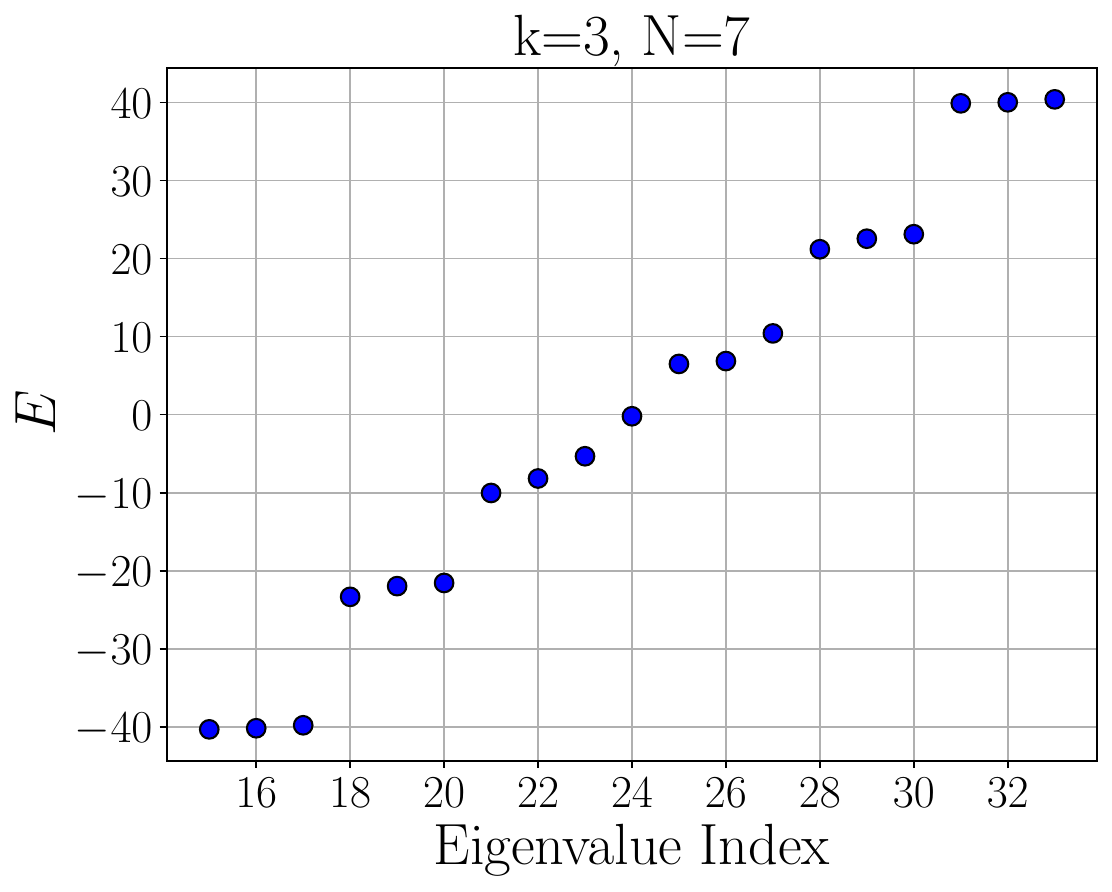}
    \caption{}
    \end{subfigure}
    \begin{subfigure}{.48\linewidth}
    \includegraphics[width=1\linewidth]{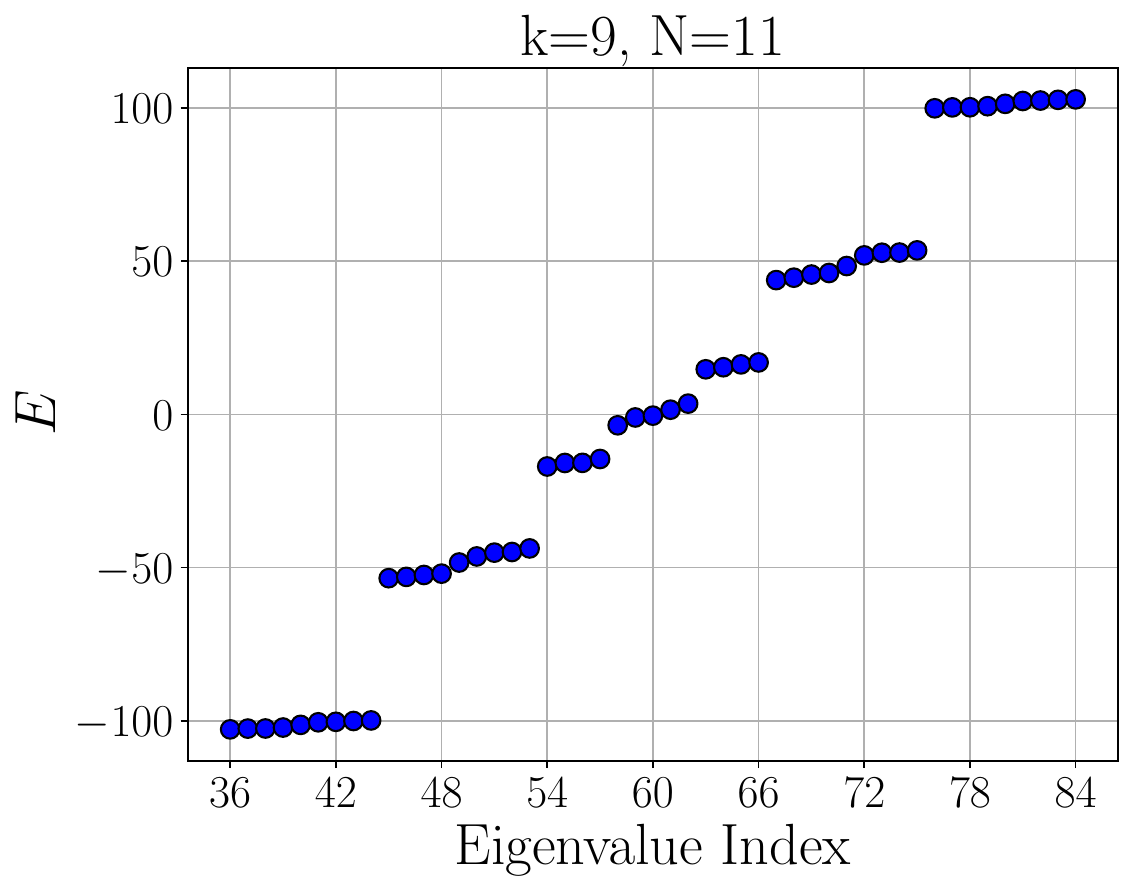}
    \caption{}
    \end{subfigure}
    \caption{Mid spectrum for (a) $k=3$, $N = 7$ and (b) $k=9$, $N = 11$.}
    \label{fig:mid_spectrum_k_generic_N_odd}
\end{figure}

When $\gcd(k,N)=1$, the action of the twisted boundary conditions generates a single orbit spanning all Fourier sectors. Indeed, the map
\begin{equation}
n \longrightarrow n+k \pmod N
\end{equation}
generates one orbit of length $N$:
\begin{equation}
N_{\mathrm{orb}}=1,
\qquad
\ell_{\mathrm{orb}}=N.
\end{equation}
As a consequence, all Fourier sectors are globally connected under a winding in the $x$ direction, and the Hilbert space does not decompose into independent invariant subsectors.

This has important consequences for the spectral structure. In particular, the continuum $k$-fold degeneracy cannot arise from an exact discrete symmetry algebra acting separately on disconnected sectors, as happens in the commensurate case $k\mid N$, as we will see in the next section. Instead, all sectors are mixed together into a single irreducible structure, and the emergence of quasi-degenerate $k$-plets becomes a purely dynamical effect associated with the continuum limit $\alpha\to0$.
The spectrum generically consists of quasi-degenerate $k$-plets together with a small number of unpaired levels.

The origin of the latter can be understood from simple algebraic considerations. We distinguish two qualitatively different cases:
\begin{enumerate}
    \item $N$ even and $k$ odd,
    \item $N$ odd for generic $k$.
\end{enumerate}

The first of such cases is an explicit example of how chiral symmetry organizes the spectrum. For every $k$-fold multiplet with energy $E$, chiral symmetry requires the existence of a corresponding multiplet at energy $-E$. The spectrum therefore maximizes the number of such paired multiplets within the available Hilbert-space dimension, $N^2$. We exemplify this behavior in Fig.~\ref{fig:mid_spectrum_k_odd_N_even}. For $k=3$ and $N=4$, the spectrum contains two pairs of quasi-degenerate three-fold multiplets. Although enough states remain to form an additional multiplet, such a multiplet could not be paired with a chiral partner and therefore does not occur. Instead, the remaining four states remain ungrouped. The same happens for $k=5$ and $N=6$, where the spectrum hosts three pairs of multiplets in total (for visualization purposes only two pairs are shown in Fig.~\ref{fig:mid_spectrum_k_odd_N_even}) and additional six unpaired states in the middle of the spectrum.

For odd $N$ chiral symmetry is explicitly broken. Nevertheless, numerical results indicate that an approximate chiral symmetry emerges for sufficiently large $N$. Consequently, the spectrum exhibits a structure similar to that of the even-$N$ case. For example for $k=3$, $N=7$ (Fig.~\ref{fig:mid_spectrum_k_generic_N_odd} left panel) the spectrum consists of $8$ pairs of multiplets in total, whose degeneracy becomes strongly broken towards $E=0$, and an isolated energy level close to (but not exactly at) $E=0$. For $k=9$, $N=11$, we observe a spectrum of $6$ pairs of nine-fold quasi-degenerate multiplets. Although the absence of an exact chiral symmetry would in principle allow an additional multiplet, no such structure is present: rather, the remaining $13$ states are not organized in a $k$-fold structure. The detailed arrangement of these unpaired mid-spectrum states appears to be dynamical rather than algebraic in origin, and we do not currently have a complete explanation for their organization.

Such apparent mismatch with the continuum $k$-fold degeneracy is resolved by taking the limit
\begin{equation}
N \to \infty,
\qquad
k \ \text{fixed},
\qquad
\alpha = \frac{k}{N^2} \to 0.
\end{equation}
In this limit:
\begin{itemize}
\item the number of quasi-degenerate $k$-plets grows as $\sim N^2/k$,
\item the splittings within each multiplet vanish,
\item the relative weights of the mid-spectrum states tend to zero,
\begin{equation}
\frac{1}{N^2} \longrightarrow 0.
\end{equation}
\end{itemize}
Thus, the continuum $k$-fold degeneracy is recovered in the bulk spectrum, while finite-size states become irrelevant in the continuum limit.
This is also related to the question concerning why the breakdown of the $k$-fold structure is observed specifically near $E=0$. In particular, recall that for a Hamiltonian $H$ with a finite spectrum, the highest-energy states of $H$ correspond to the low-energy states of $-H$. They are therefore expected to exhibit the same characteristic low-energy features. In the present case, this suggests that, even for odd $N$, where the spectrum is not exactly symmetric, the highest-energy states should still organize into quasi-degenerate $k$-plets.
Consequently, the portion of the spectrum that is more likely to be sensitive to lattice artefacts is the central region, where we indeed observe effects related to the incommensurability between the finite Hilbert space and the approximate $k$-fold degeneracy.

\subsubsection{Case II: $k \mid N$}
\label{subsec:case_II}

\begin{figure}
    \centering
    \begin{subfigure}{.48\linewidth}
    \includegraphics[width=1\linewidth]{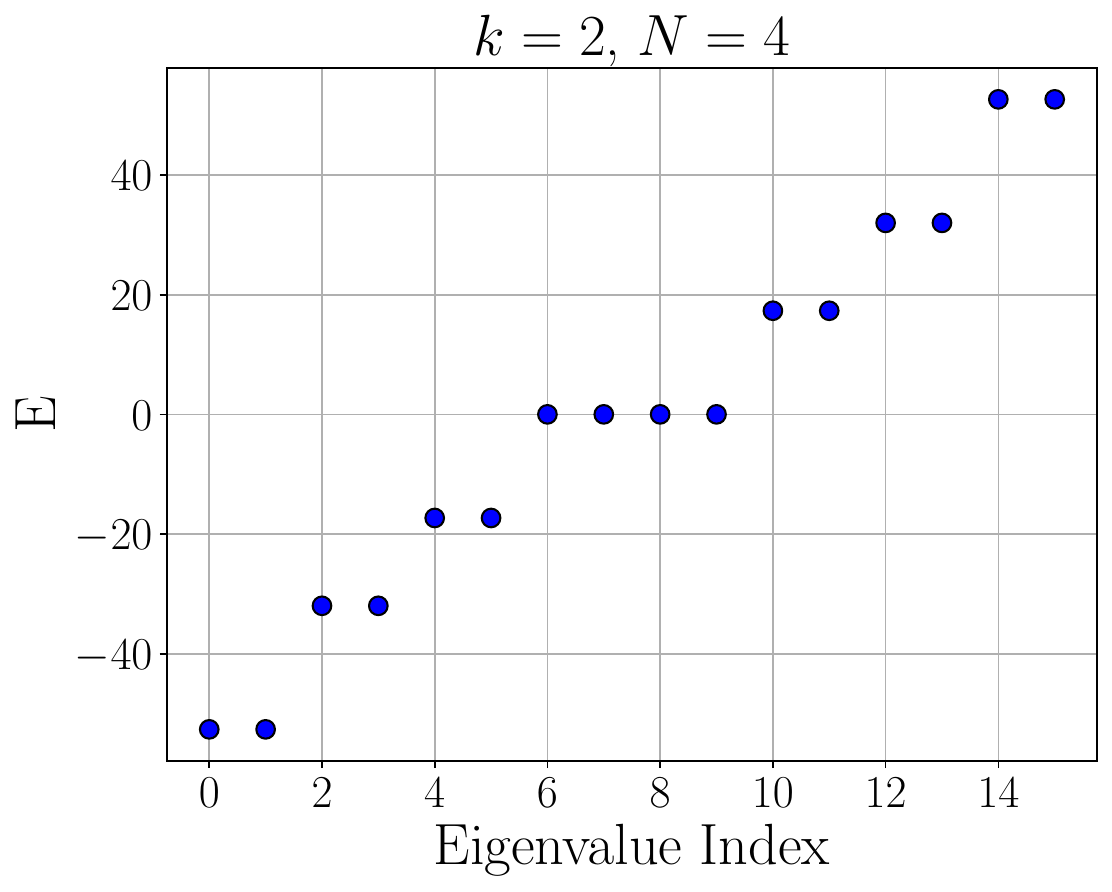}
    \caption{}
    \end{subfigure}
    \begin{subfigure}{.48\linewidth}
    \includegraphics[width=1\linewidth]{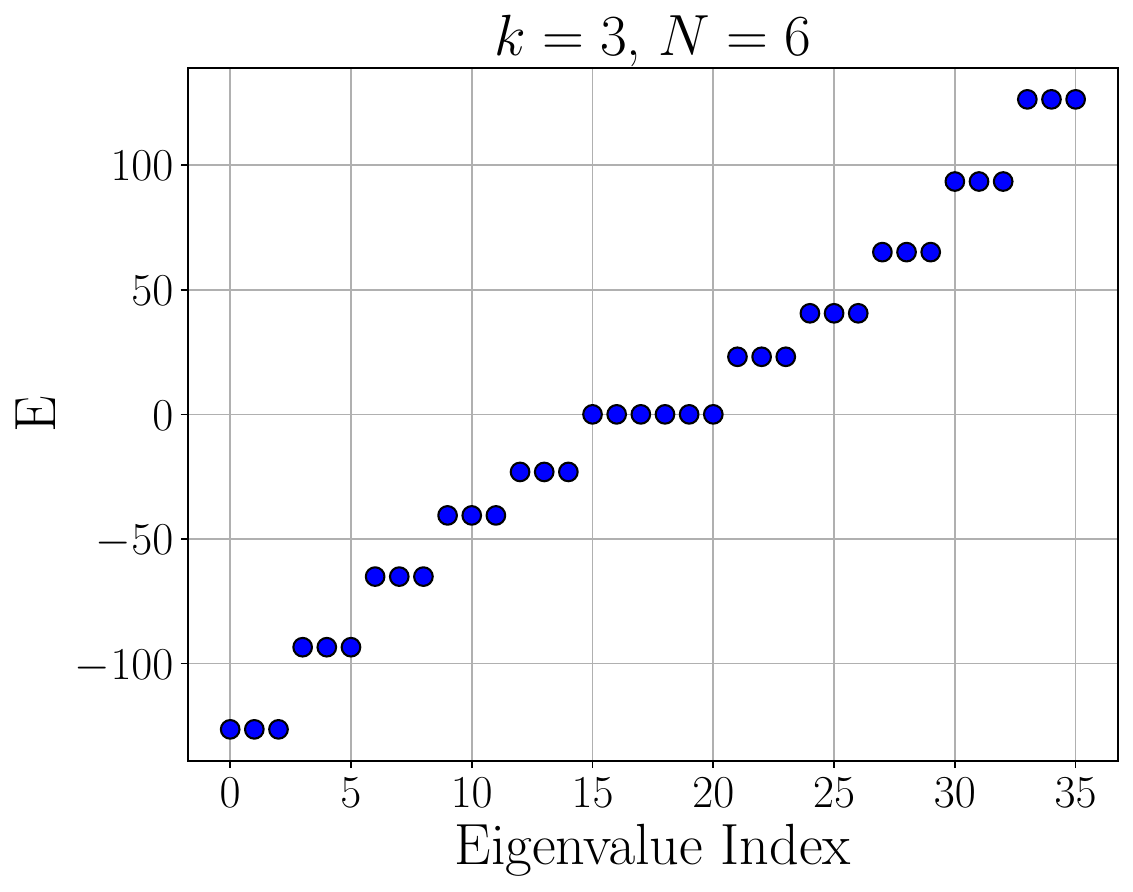}
    \caption{}
    \end{subfigure}
    \caption{Full spectrum for (a) $k=2$, $N = 4$ and (b) $k=3$, $N = 6$.}
    \label{fig:full_spectrum_case_II}
\end{figure}

We now consider the commensurate case
\begin{equation}
N = kr.
\end{equation}
In this situation,
\begin{equation}
\gcd(k,N)=k,
\end{equation}
and therefore
\begin{equation}
N_{\text{orb}}=k,
\qquad
\ell_{\text{orb}}=\frac{N}{k}=r.
\end{equation}
Thus, the Fourier sectors decompose into $k$ disjoint orbits, each of length $r$.

This is the cleanest arithmetic regime of the discretized model. The spectrum naturally organizes into $k$ families, and the low-energy region displays an especially transparent $k$-plet structure. In this regime, the lattice realizes the continuum topological pattern in the most direct way. An example is shown in Fig.~\ref{fig:full_spectrum_case_II} for $k=2$, $N=4$ and for $k=3$, $N=6$.

The exact $k$-fold degeneracy originates from a reduced magnetic translation symmetry that emerges when the commensurability conditions
\begin{equation}
k\mid N_x,
\qquad
k\mid N_y,
\label{comcon}
\end{equation}
are satisfied. For the square lattice case considered here, this reduces to $k\mid N$.

Setting
\begin{equation}
m_x=\frac{N_x}{k},
\qquad
m_y=\frac{N_y}{k},
\label{gcd}
\end{equation}
we introduce the reduced magnetic translation operators
\begin{equation}
X=(T_x)^{m_x},
\qquad
Z=(T_y)^{m_y}.
\end{equation}
As shown in Appendix~\ref{app:magnetic-translations-details}, these operators commute with the Hamiltonian,
\begin{equation}
[X,H]=0,
\qquad
[Z,H]=0,
\label{comtot}
\end{equation}
both in the bulk and at the boundary links.

Starting from the elementary magnetic translation algebra
\begin{equation}
T_xT_y=e^{i2\pi\alpha}T_yT_x,
\end{equation}
one finds
\begin{align}
XZ
&=
T_x^{m_x}T_y^{m_y}
=
e^{i2\pi\alpha m_xm_y}\,
T_y^{m_y}T_x^{m_x}
=
e^{i2\pi/k}\,ZX,
\end{align}
since
\begin{equation}
\alpha m_xm_y
=
\frac{k}{N_xN_y}
\frac{N_x}{k}
\frac{N_y}{k}
=
\frac{1}{k}.
\end{equation}
Therefore,
\begin{equation}
XZ=e^{i2\pi/k}ZX.
\label{eq:projective-final}
\end{equation}

The operators $X$ and $Z$ thus realize a finite Weyl algebra, which is the lattice counterpart of the algebra governing topological degeneracy in the continuum theory. 

The existence of exact $k$-fold multiplets follows immediately. Let $\psi$ be a simultaneous eigenstate of $H$ and $Z$,
\begin{equation}
H\psi = E\psi,
\qquad
Z\psi = z\psi.
\end{equation}
Since $[H,X]=0$, the states
\begin{equation}
\psi,\quad X\psi,\quad X^2\psi,\quad \dots,\quad X^{k-1}\psi
\end{equation}
all have the same energy $E$. Using Eq.~\eqref{eq:projective-final}, one obtains
\begin{equation}
Z(X^n\psi)
=e^{-i2\pi n/k}\,
z\,X^n\psi,
\qquad
n=0,\dots,k-1.
\end{equation}
The states, therefore, carry distinct $Z$ eigenvalues and are linearly independent, forming an exact $k$-dimensional multiplet.

In the commensurate case, the boundary twist becomes compatible with the reduced translations, even though it obstructs the global periodicity of the elementary magnetic translations. The reduced operators $X$ and $Z$ therefore provide an exact lattice realization of the continuum topological algebra.

In Fig.~\ref{fig:full_spectrum_case_II}, one observes that for both $k=2$, $N=4$ and $k=3$, $N=6$ the spectrum contains $2k$ degenerate states at $E=0$, rather than $k$. This enhancement is a consequence of the additional chiral symmetry present for even lattice sizes. Since the spectrum is symmetric under $E \longrightarrow -E$, the two $k=2$ and $k=3$ multiplets related by chiral symmetry coincide at zero energy, producing an accidental enhanced degeneracy.

Note that if only one of the two conditions in Eq.~\eqref{comcon} is satisfied, i.e., if the lattice is commensurate only along one direction, the pair of reduced magnetic translations generating the projective algebra Eq.~\eqref{eq:projective-final} cannot be constructed. As a consequence, the finite Weyl algebra responsible for the exact $k$-fold degeneracy is not realized, and the exact degeneracy is lost. This case will correspond to a rectangular discretization ($N_x\neq N_y$). Then, the lattice Hamiltonian leads to an anisotropic Harper equation characterized by different hopping amplitudes along the two directions.
For a rectangular discretization, one has
\begin{equation}
\Delta_x=\frac{2\pi}{N_x},
\qquad
\Delta_y=\frac{2\pi}{N_y}.
\end{equation}
Therefore, the hopping amplitudes are different:
\begin{equation}
t_x=\frac{e^2 S}{2 \Delta_x^2},
\qquad
t_y=\frac{e^2 S}{2 \Delta_y^2}.
\end{equation}
  The resulting Harper equation is
anisotropic:
\begin{equation}
E g_n(x)
=
-t_x[g_n(x+1)+g_n(x-1)]
-2t_y\cos(2\pi\alpha x+p_y)g_n(x) ,
\qquad
\alpha=\frac{k}{N_xN_y}.
\end{equation}

\begin{figure}
    \centering
    \begin{subfigure}{.48\linewidth}
    \includegraphics[width=1\linewidth]{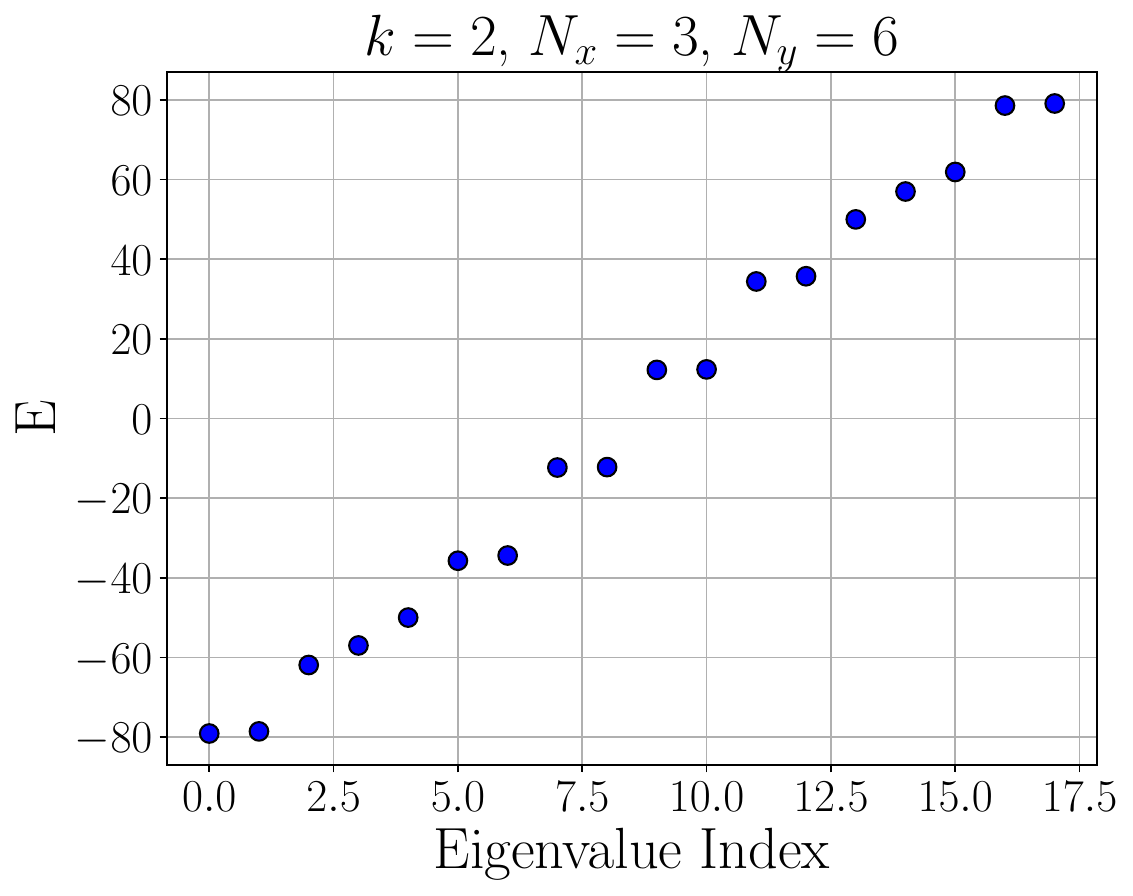}
    \caption{}
    \end{subfigure}
    \begin{subfigure}{.48\linewidth}
    \includegraphics[width=1\linewidth]{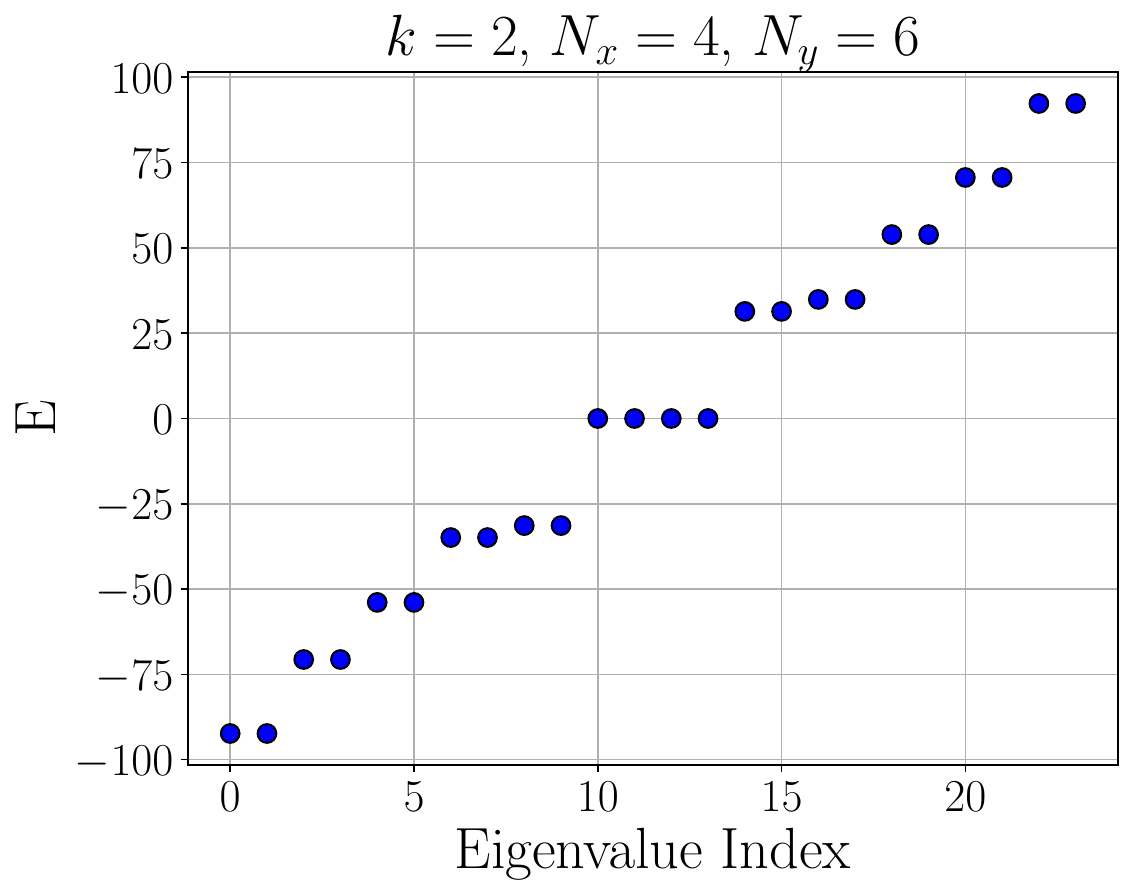}
    \caption{}
    \end{subfigure}
    \caption{Full spectrum for (a) $k=2$, $N_x = 3$ and $N_y = 6$ and (b) $k=2$, $N_x = 4$ and $N_y=6$.}
    \label{fig:asymmetric_case}
\end{figure}

We have verified numerically that this anisotropy modifies the energy spectrum but does not affect the underlying magnetic-translation algebra. In particular, whenever the commensurability conditions (\ref{comcon}) are satisfied, the exact topological degeneracy remains protected despite the anisotropy as shown in Fig.~\ref{fig:asymmetric_case}, right panel, for $k=2,\  N_x=4,\  N_y=6$.

An example, instead, when commensurability is broken in one direction is shown in Fig.~\ref{fig:asymmetric_case}, left panel, for $k=2$, $N_x=3$, and $N_y=6$.

\subsection{Discretizations outside the continuum scaling regime}
\label{sec:other_regimes}

The cases discussed in this section do not belong to the continuum scaling regime of the MCS zero-mode sector. Although they define perfectly consistent finite-dimensional lattice systems, the discretization is too coarse to faithfully resolve the continuum $k$-fold topological structure. Their spectra are instead dominated by arithmetic commensurability effects, twisted boundary conditions, and finite-size Hofstadter-like physics. Nevertheless, these regimes exhibit interesting algebraic and spectral structures, which are worth analyzing in their own right.

\subsubsection{Case III: $\gcd(k,N)=1$ with $k>N$}

\begin{figure}
    \centering
    \includegraphics[width=0.48\linewidth]{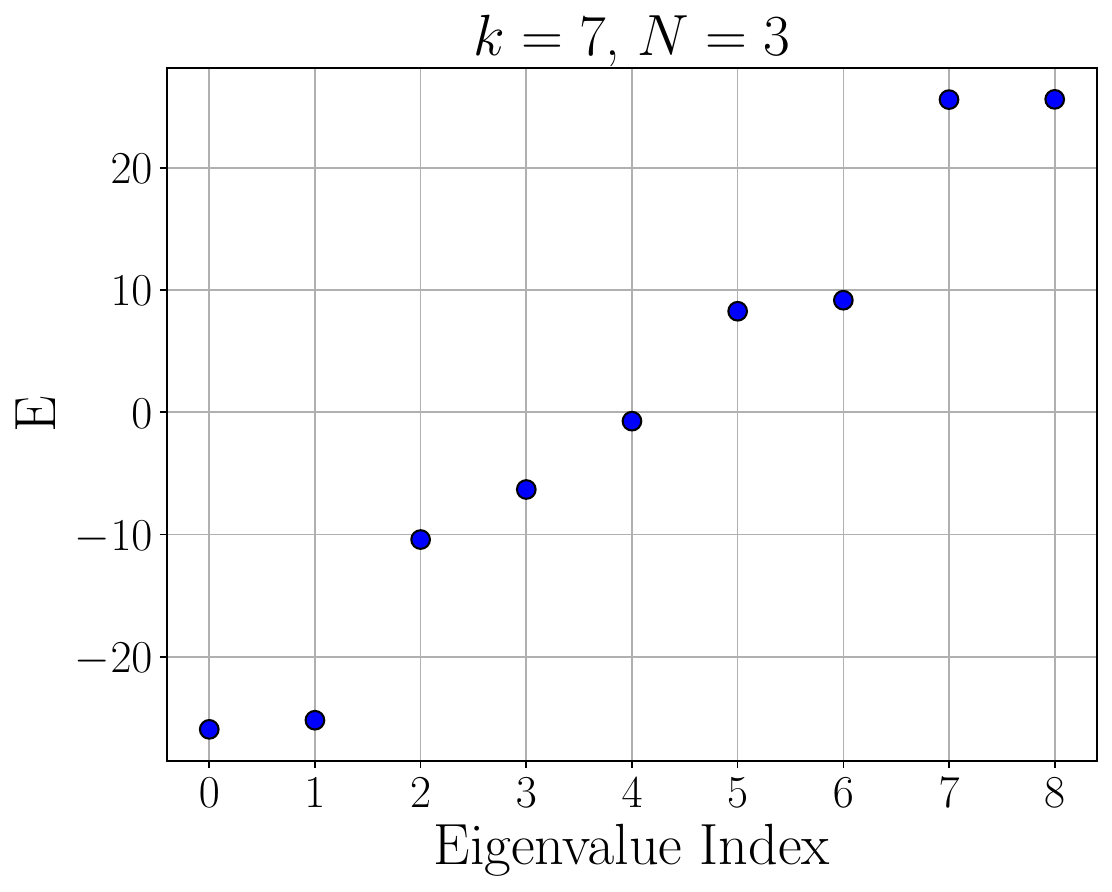}
    \caption{Full spectrum for $k=7$, $N = 3$.}
    \label{fig:full_spectrum_k7_N3}
\end{figure}

We now reconsider the case
\begin{equation}
\gcd(k,N)=1,
\qquad
\alpha<1,
\label{caseIII}
\end{equation}
but with
\begin{equation}
k>N, \qquad k< N^2,
\end{equation}
which  requires a refinement of the previous discussion.

Since $k$ and $N^2$ are generically coprime, the flux per plaquette is already in lowest terms,
\begin{equation}
\alpha=\frac{k}{N^2} = \frac{p}{q},
\qquad
q=N^2.
\end{equation}

In the standard Hofstadter problem, a rational flux $\alpha=p/q$ is associated with a magnetic periodicity of size $q$~\cite{Fradkin:2013anc}. In the present case,
\begin{equation}
q=N^2>N,
\end{equation}
so the corresponding magnetic scale is larger than the linear size of the lattice. In particular, one cannot construct a magnetic unit cell inside the finite torus.

As discussed previously, when $\gcd(k,N)=1$ the twisted boundary conditions generate a single orbit involving all Fourier sectors. This remains true independently of whether $k<N$ or $k>N$. The crucial difference between the two regimes is therefore not the orbit structure itself, but the ability of the lattice to resolve the semiclassical guiding centers associated with the continuum Landau problem.

For $k<N$, one has
\begin{equation}
\frac{N}{k}\gg1,
\end{equation}
so that the guiding centers are well separated on the lattice and the spectrum organizes into quasi-degenerate $k$-plets approaching the continuum Landau-level structure.
By contrast, when $k>N$, although the orbit structure is unchanged, the discretization becomes too coarse to resolve the individual guiding centers. In this regime, the global mixing of Fourier sectors induced by the twist dominates the spectrum, and the continuum $k$-fold structure is no longer visible.

Because of the absence of commensurability and the maximal mixing induced by the twist, the spectrum does not organize into well-defined Hofstadter sub-bands.  Instead, one obtains a fully discrete spectrum with $N^2$ levels, whose structure reflects a nontrivial interplay between:
\begin{itemize}
\item the rational value of $\alpha$,
\item the twisted boundary conditions,
\item the finite-dimensional Hilbert space.
\end{itemize}
An example is shown in Fig.~\ref{fig:full_spectrum_k7_N3} for
\begin{equation}
k=7,
\qquad
N=3,
\qquad
\alpha=\frac{7}{9}.
\end{equation}
In this case,
\begin{equation}
n\mapsto n+7 \equiv n+1 \pmod 3,
\end{equation}
so all Fourier sectors belong to a single orbit. The resulting spectrum consists of nine discrete levels with no clear quasi-degenerate multiplet structure.

\begin{figure}
    \centering
    \includegraphics[width=0.48\linewidth]{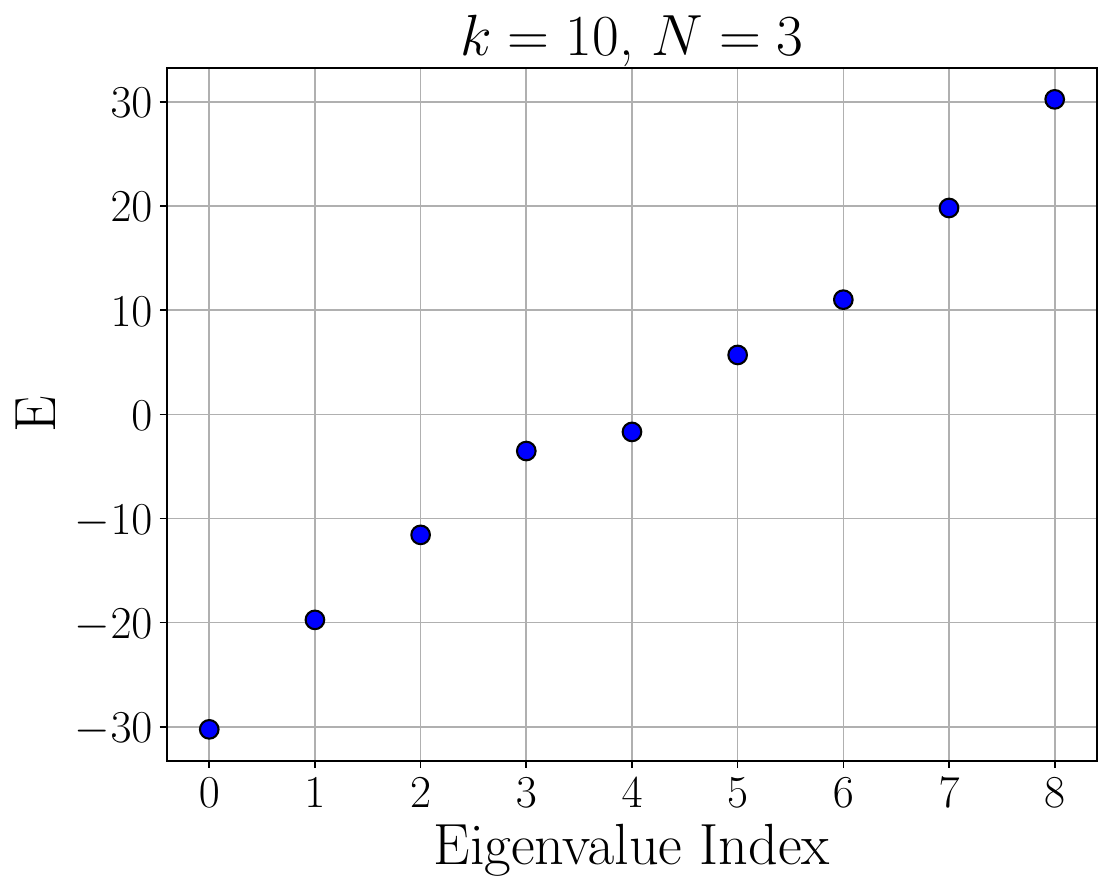}
    \caption{Full spectrum for $k=10$, $N = 3$.}
    \label{fig:full_spectrum_k10_N3}
\end{figure}

We now consider the regime
\begin{equation}
k>N^2,
\qquad
\gcd(k,N)=1,
\label{casoIIIalpha}
\end{equation}
so that
\begin{equation}
\alpha=\frac{k}{N^2}>1.
\end{equation}
Since the lattice phases are of the form
$e^{i2\pi\alpha}$,
only the fractional part of $\alpha$ is physically relevant for the local magnetic structure\footnote{Notice that this also implies that the spectrum for $N^2 < k < 2N^2$ is the same as the spectrum for $2N^2 < k < 3N^2$ and so on.}. One can therefore write
\begin{equation}
\alpha
=
\frac{k}{N^2}
=
1+\frac{k-N^2}{N^2}
=
1+\frac{p}{q},
\end{equation}
with
\begin{equation}
p=k-N^2,
\qquad
q=N^2.
\end{equation}
Using the identity
\begin{equation}
\gcd(k-N^2,N^2)=\gcd(k,N^2),
\end{equation}
together with $\gcd(k,N)=1$, one finds
\begin{equation}
\gcd(p,q)=1.
\end{equation}
Thus the effective flux
\begin{equation}
\alpha_{\rm eff}
=
\frac{k-N^2}{N^2}
\end{equation}
is already in lowest terms.

In this regime, the effective magnetic periodicity becomes comparable to the full discretized torus itself. The system therefore exhibits a Hofstadter-like commensurability structure controlled by the fractional part of the flux rather than by the original value of $k$.

An example is provided by
\begin{equation}
k=10,
\qquad
N=3,
\qquad
\alpha=\frac{10}{9}
=
1+\frac19 \ ,
\end{equation}
shown in Fig.~\ref{fig:full_spectrum_k10_N3}.
In this case, the effective local flux is
\begin{equation}
\alpha_{\rm eff}=\frac19.
\end{equation}
Thus, even for very large values of $k$, the finite discretized torus is insensitive to the integer part of $\alpha$. The spectral structure is instead determined by the effective local flux per plaquette together with the twisted boundary conditions and the finite-dimensional geometry of the zero-mode Hilbert space.

\subsubsection{Case IV: $N \mid k$}

\begin{figure}
    \centering
    \includegraphics[width=0.48\linewidth]{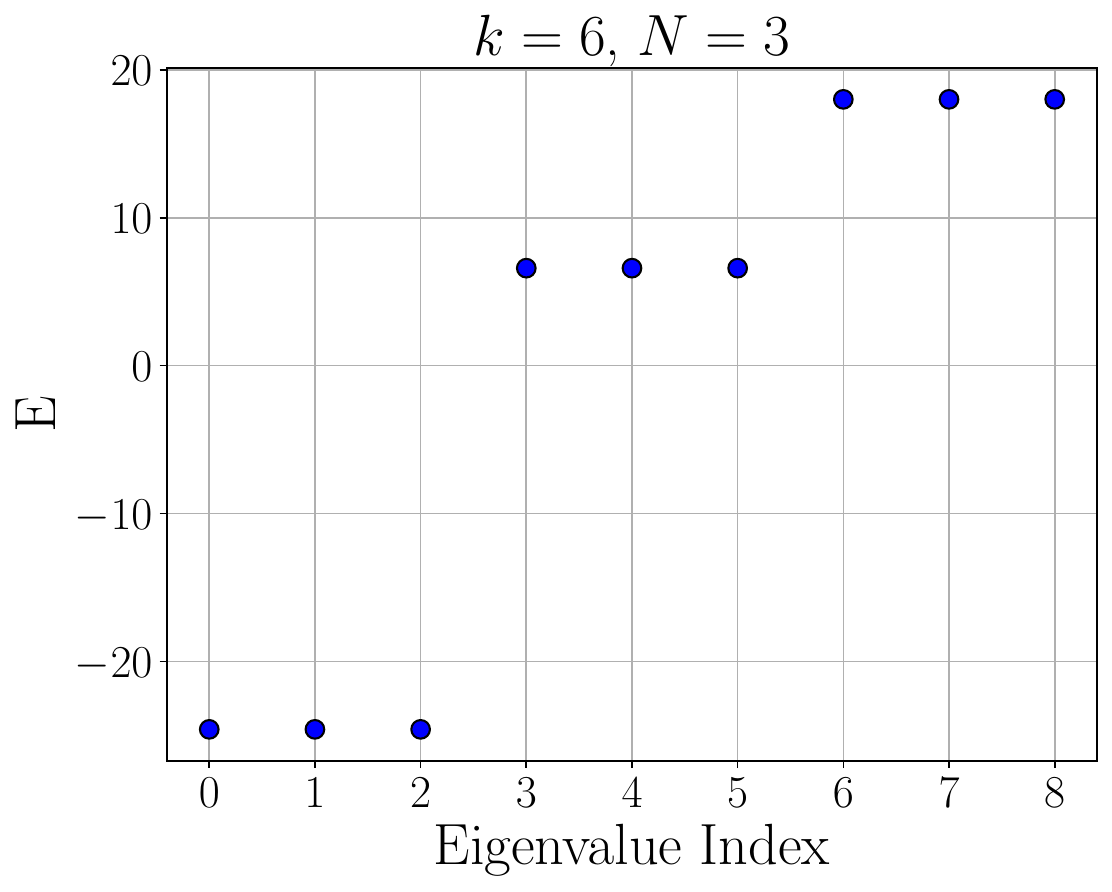}
    \caption{Full spectrum for $k=6$, $N=3$.}
    \label{fig:full_spectrum_k6_N3}
\end{figure}

We now discuss the case in which $k$ is an integer multiple of $N$. In this regime, the boundary twist becomes trivial since
\begin{equation}
\frac{k}{N}\in\mathbb Z,
\end{equation}
and the lattice contains at least one magnetic unit cell. These configurations do not belong to the continuum scaling regime of the zero-mode sector, since the continuum limit requires
$N\to\infty$ and $k$ fixed.
Their spectral structure is therefore governed by Hofstadter-type lattice physics rather than by the continuum $k$-fold topological degeneracy.

We first consider the case
\begin{equation}
k=mN,
\qquad
m<N,
\qquad
\alpha=\frac{k}{N^2}=\frac{m}{N}<1.
\label{caso41}
\end{equation}
In this regime, the system reduces to a standard Hofstadter problem with rational flux
\begin{equation}
\alpha=\frac{m}{N}.
\end{equation}
The effective flux is in lowest terms only if $(m,N)=1$. In general,
\begin{equation}
\alpha=\frac{p}{q},
\qquad
q=\frac{N}{\gcd(m,N)}.
\end{equation}
The spectrum then organizes into $q$ groups, each containing
\begin{equation}
\frac{N^2}{q}
\end{equation}
states.
An example is shown in Fig.~\ref{fig:full_spectrum_k6_N3} for
\begin{equation}
k=6,
\qquad
N=3,
\qquad
\alpha=\frac{2}{3}.
\end{equation}

We next consider the case
\begin{equation}
k=mN,
\qquad
m\ge N,
\end{equation}
for which
\begin{equation}
\alpha=\frac{m}{N}\ge1.
\end{equation}
Writing
\begin{equation}
m=rN+s,
\qquad
0\le s < N,
\end{equation}
one obtains
\begin{equation}
\alpha
=r+\frac{s}{N}.
\end{equation}
Reducing the fractional part to lowest terms gives
\begin{equation}
\alpha_{\rm eff}
=\frac{p}{q},
\qquad
p=\frac{s}{\gcd(s,N)},
\qquad
q=\frac{N}{\gcd(s,N)}.
\end{equation}
Also in this case, the spectrum is therefore controlled by the effective fractional flux $\alpha_{\rm eff}$, and it organizes into $q$ groups of size
\begin{equation}
\frac{N^2}{q}.
\end{equation}
\begin{figure}
    \centering
    \includegraphics[width=0.48\linewidth]{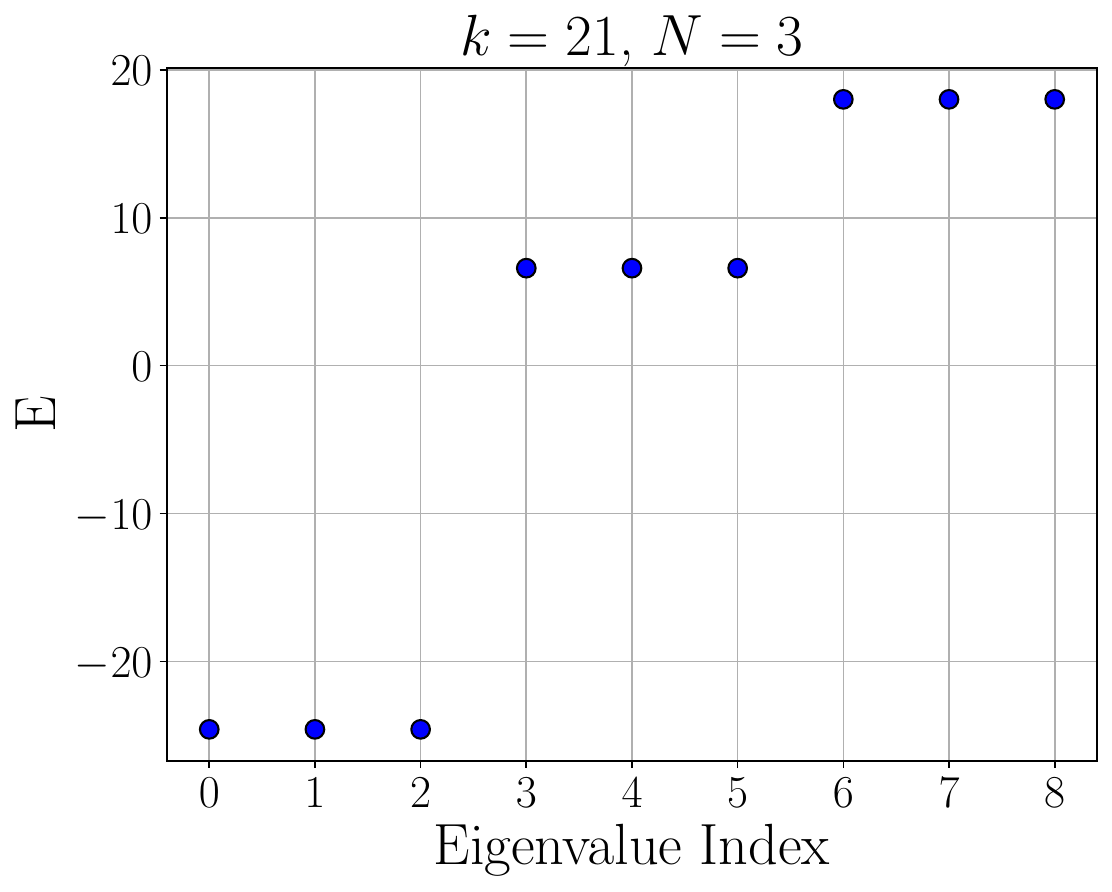}
    \caption{Full spectrum for $k=21$, $N = 3$.}
    \label{fig:full_spectrum_k21_N3}
\end{figure}
An example is shown in Fig.~\ref{fig:full_spectrum_k21_N3} for
\begin{equation}
k=21,
\qquad
N=3,
\qquad
\alpha=\frac{21}{9}=2+\frac13.
\end{equation}
In this case,
\begin{equation}
\alpha_{\rm eff}=\frac{1}{3},
\end{equation}
so the discretized zero-mode problem becomes equivalent to a Hofstadter model with flux $1/3$ on a finite $3\times3$ torus.

The Hilbert space has dimension
\begin{equation}
\dim\mathcal H=N^2=9,
\end{equation}
and the spectrum therefore organizes into
\begin{equation}
q=3
\end{equation}
multiplets containing
\begin{equation}
\frac{N^2}{q}=3
\end{equation}
states each. This explains why the observed degeneracy is $3$, rather than the continuum value $k=21$.

A particularly simple situation occurs when
\begin{equation}
N^2\mid k,
\end{equation}
for which
\begin{equation}
\alpha=\frac{k}{N^2}\in\mathbb Z.
\end{equation}
In this case,
\begin{equation}
U_y(x)=e^{2\pi i\alpha x}=1
\end{equation}
for every lattice site $x$, and the Hofstadter Hamiltonian reduces exactly to the standard nearest-neighbor tight-binding Hamiltonian on a periodic square lattice,
\begin{equation}
H=-t\sum_{x,y}
\Bigl(
|x,y\rangle\langle x+1,y|
+
|x,y\rangle\langle x,y+1|
+\mathrm{h.c.}
\Bigr).
\label{tbh}
\end{equation}
The magnetic structure therefore disappears completely, and any remaining degeneracies are due solely to lattice symmetries and momentum quantization. In this regime, the lattice no longer resolves the topological information associated with the continuum flux $2\pi k$.

\begin{figure}
    \centering
    \begin{subfigure}{.32\linewidth}
    \includegraphics[width=1\linewidth]{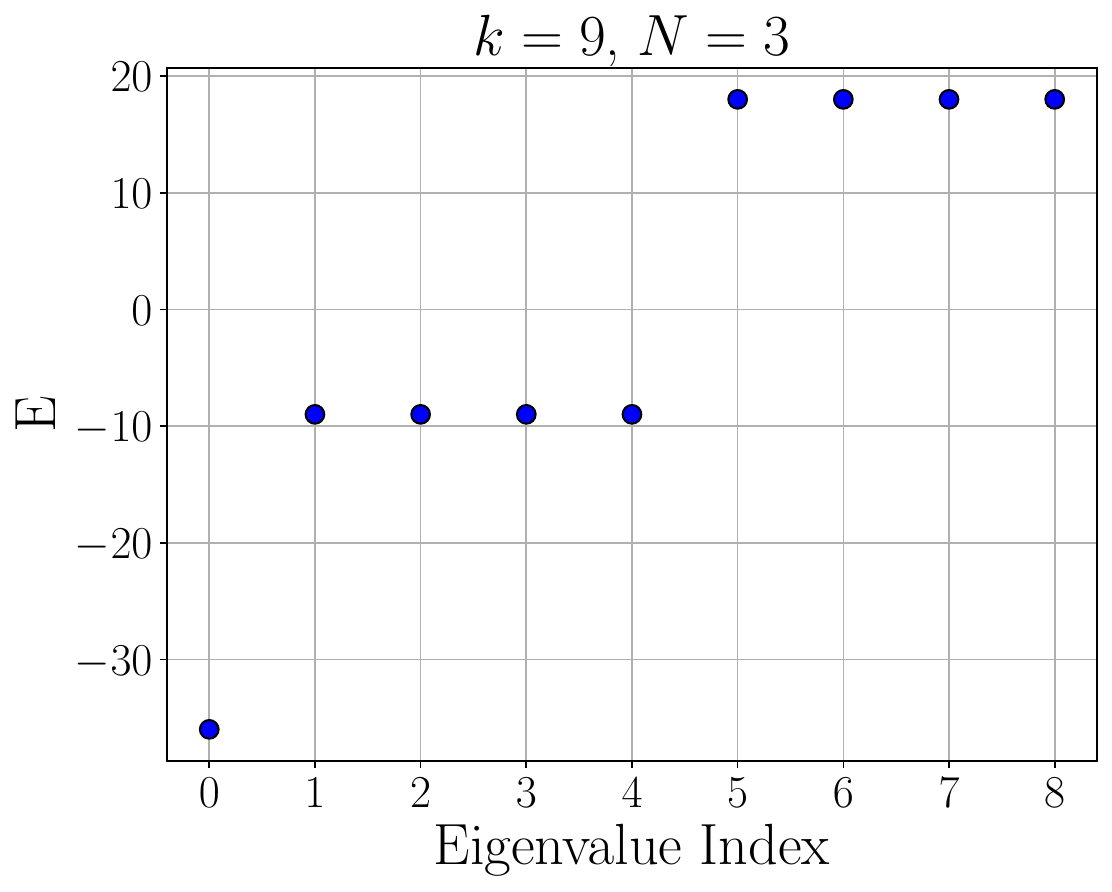}
    \caption{}
    \end{subfigure}
    \begin{subfigure}{.32\linewidth}
    \includegraphics[width=1\linewidth]{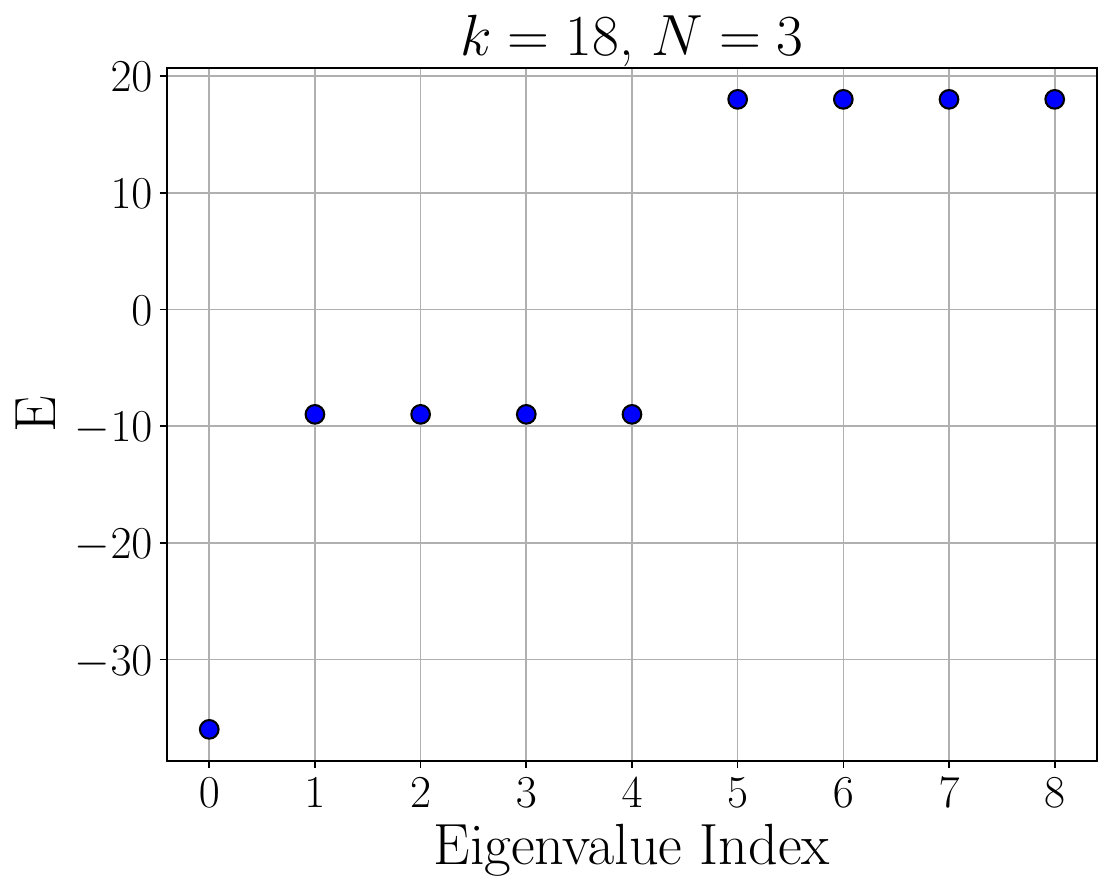}
    \caption{}
    \end{subfigure}
    \begin{subfigure}{.32\linewidth}
    \includegraphics[width=1\linewidth]{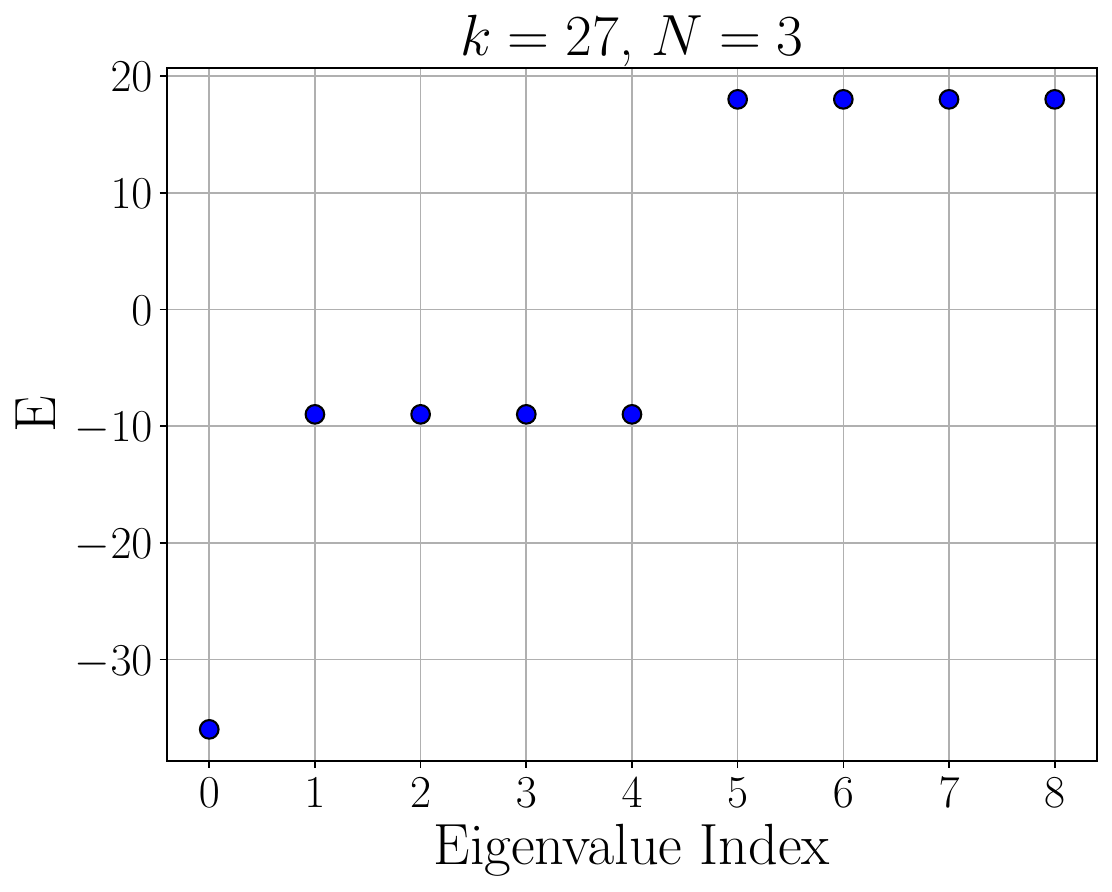}
    \caption{}
    \end{subfigure}
    \caption{Full spectrum for (a) $k=9$ (b) $k=18$ and (c) $k=27$ and $N=3$.}
    \label{fig:full_spectrum_case_k9_18_27}
\end{figure}

For example, on a $3\times3$ lattice, all values
\begin{equation}
k=9,18,27,\dots
\end{equation}
lead to the same zero-field spectrum, as shown in Fig.~\ref{fig:full_spectrum_case_k9_18_27}.

\subsubsection{Hofstadter butterfly}

Finally, we analyze the spectrum as the flux per plaquette, $\alpha$, is varied at fixed $N$. Within our construction, changing $\alpha$ at fixed truncation corresponds to varying the Chern--Simons level while keeping the local Hilbert-space dimension unchanged. Although this procedure does not correspond to approaching the continuum limit of a fixed continuum theory, it provides a useful probe of the microscopic lattice model.

\begin{figure}
    \centering
    \includegraphics[width=0.6\linewidth]{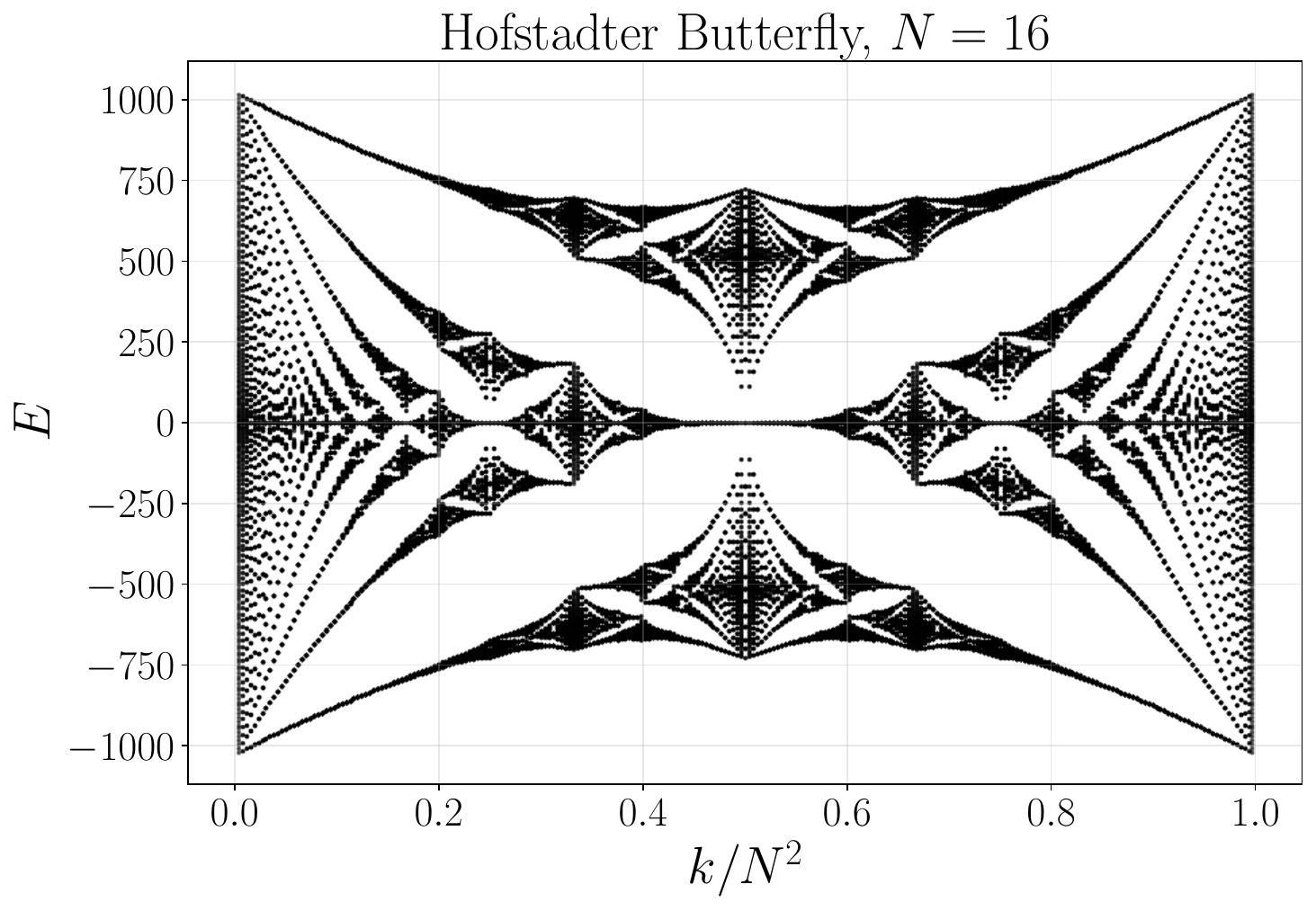}
    \caption{Energy spectrum as a function of the flux per plaquette at fixed $N$.}
    \label{fig:butterfly}
\end{figure}

The resulting spectrum for $N=16$ is shown in Fig.~\ref{fig:butterfly}. As $\alpha$ is varied, the energy levels reorganize into a complex hierarchy of bands and gaps. The figure clearly exhibits the characteristic self-similar structure known as the Hofstadter butterfly~\cite{Hofstadter:1976}, originally derived for electrons hopping on a lattice in the presence of a magnetic field. From Fig.~\ref{fig:butterfly}, the symmetry upon replacing $k/N^2$ with $1-k/N^2$ becomes evident; this is a consequence of the cosine term being periodic in the Harper equation.

The emergence of the Hofstadter butterfly provides a nontrivial consistency check of the lattice formulation. Starting from a continuum Chern--Simons theory (regularized with a Maxwell term), which is known to provide a low-energy effective description of the quantum Hall effect, the discretization and truncation procedures provide precisely a model of an electron in a background magnetic field.

\section{Conclusions}
\label{sec:conclusions}

In this work, we have studied the Hilbert space truncation of the constant modes in the Hamiltonian MCS theory. With the eventual goal of simulating nonperturbative features of the full theory by means of Hamiltonian simulations, the constant-mode sector of the theory provides a controlled setup for testing truncation schemes. Our approach consists of discretizing the target-space manifold of the constant modes, namely, a torus. The discretization of the lattice can be controlled by varying the linear size of the torus $N$, with the continuum limit corresponding to $N\rightarrow\infty$.

Within this scheme, we were able to map the Hamiltonian of the constant modes into a particle hopping on a periodic lattice subject to a magnetic field: this framework is equivalent to a Hofstadter problem with twisted boundary conditions. The peculiarity of our construction with respect to a standard Hofstadter problem is the fact that the flux per plaquette, $\alpha = k/N_xN_y$, is determined by $k$, a physical datum from the continuum MCS theory, and the size of the lattice, induced by the truncation scheme. Therefore, unlike the standard problem, $\alpha$ is not a physical quantity that can be adjusted independently of the lattice size. As a consequence, the physics of our model qualitatively differs from that of a standard Hofstadter problem.

The associated Harper equation can be solved by exact diagonalization, enabling a systematic analysis of the spectral properties at different values of $k$ and $N$. We investigate several regimes. For $k > N$, the system exhibits Hofstadter-like behavior, and the topological $k$-fold degeneracy present in continuum theory is lost. In contrast, this degeneracy is asymptotically recovered in the regime $k < N$. A systematic study of the continuum limit shows that the degeneracy is restored exponentially fast with increasing $N$, whereas the spectrum converges algebraically to that of a harmonic oscillator.

Even though the discretization and simulation of the full theory remain an open problem that we are planning to investigate in future work, our work provides a first study of the effects induced by truncating the spectrum of the MCS theory. Interestingly, key physical properties are already visible at coarse truncations and exhibit clear convergence towards the continuum limit.

\appendix

\section{Explicit derivation of the lattice Hamiltonian from the continuum}
\label{app:lattice-hamiltonian-details}
In this appendix we show the detailed steps to arrive at the Hamiltonian~\eqref{hfm1}.
We start off by inserting Eq.~\eqref{sym} into Eq.~\eqref{hfm}:
\begin{align*}
    H &=\frac{S e^2}{2}\left[\left(p_1 + \tilde{A}_1\right)^2 + \left(p_2 +\tilde{A}_2\right)^2\right]\\
    &=\frac{S e^2}{2}\left[\left(-i\partial_{a_1} + \tilde{A}_1\right)^2 + \left(-i\partial_{a_2} +\tilde{A}_2\right)^2\right]\\
    &=\frac{S e^2}{2}\left[\left(-i(\partial_{a_1} + i\tilde{A}_1)\right)^2 + \left(-i(\partial_{a_2} +i\tilde{A}_2)\right)^2\right]\\
    &=\frac{S e^2}{2}\sum\limits_{i=1}^2\Pi_i^2.
\end{align*}
In the last step we introduced 
\[\Pi_j = -i(\partial_{a_j} + i\tilde{A}_j).\]
Replacing the covariant derivative by the discretized version from Eq.~\eqref{eq:cov-der+}, we arrive at
\begin{align*}
     H&= \frac{S e^2}{2}\sum\limits_{i=1}^2(-iD_i^+)^\dagger(-iD_i^+)\\
    &= \frac{S e^2}{2}\sum\limits_{i=1}^2(D_i^+)^\dagger D_i^+\\
    &= -\frac{S e^2}{2}\sum\limits_{i=1}^2D_i^- D_i^+.
\end{align*}

\section{Magnetic translations and exact degeneracy for $k\mid N_x,\;k\mid N_y$}
\label{app:magnetic-translations-details}

In this appendix, we collect the detailed algebraic checks for the magnetic
translation operators on the discretized magnetic torus. The purpose is to
make explicit which translations commute with the Harper Hamiltonian and how the exact $k$-fold degeneracy arises when the commensurability conditions
\begin{equation}
k\mid N_x,
\qquad
k\mid N_y
\end{equation}
are satisfied.

We consider a rectangular torus of size $N_x\times N_y$, with total flux
$\Phi=2\pi k$
and flux per plaquette
$\alpha=\frac{k}{N_xN_y}$.
We work in Landau gauge. The Hamiltonian density acting on the Wannier basis
$\ket{x,y}$ is
\begin{align}
h\ket{x,y}
=
-t\Big[
&\ket{x-1,y}+\ket{x+1,y}
\nonumber\\
&+e^{2\pi i\alpha x}\ket{x,y-1}
+e^{-2\pi i\alpha x}\ket{x,y+1}
\Big],
\label{eq:app-h-bulk}
\end{align}
where the diagonal term has been omitted since it gives only an overall energy
shift.

The non-trivial boundary link is the one crossing the $x$-boundary. With our
choice of convention, the patch is implemented by
\begin{equation}
\ket{x+N_x,y}
=
e^{2\pi i k y/N_y}\ket{x,y}.
\label{eq:app-twist}
\end{equation}
Equivalently, when the Hamiltonian connects $\ket{N_x-1,y}$ to $\ket{0,y}$,
the link carries the phase
\begin{equation}
U_x(N_x-1,y)=e^{2\pi i k y/N_y}.
\label{eq:app-boundary-link}
\end{equation}
With this convention,
\begin{align}
h\ket{N_x-1,y}
=
-t\Big[
&\ket{N_x-2,y}
+e^{2\pi i k y/N_y}\ket{0,y}
\nonumber\\
&+e^{2\pi i\alpha (N_x-1)}\ket{N_x-1,y-1}
\nonumber\\
&+e^{-2\pi i\alpha (N_x-1)}\ket{N_x-1,y+1}
\Big].
\label{eq:app-h-boundary}
\end{align}
Using
\begin{equation}
\alpha N_x=\frac{k}{N_y},
\end{equation}
this may also be written as
\begin{align}
h\ket{N_x-1,y}
=
-t\Big[
&\ket{N_x-2,y}
+e^{2\pi i k y/N_y}\ket{0,y}
\nonumber\\
&+e^{2\pi i k/N_y-2\pi i\alpha}\ket{N_x-1,y-1}
\nonumber\\
&+e^{-2\pi i k/N_y+2\pi i\alpha}\ket{N_x-1,y+1}
\Big].
\label{eq:app-h-boundary-expanded}
\end{align}

\subsection*{Elementary magnetic translations}

We define the elementary magnetic translations by
\begin{align}
T_x\ket{x,y}
&=
e^{-2\pi i\alpha y}\ket{x+1,y},
\label{eq:app-Tx-bulk}
\\
T_y\ket{x,y}
&=
\ket{x,y+1}.
\label{eq:app-Ty}
\end{align}
At the $x$-boundary, $T_x$ must be supplemented by the same transition
function appearing in the magnetic bundle:
\begin{equation}
T_x\ket{N_x-1,y}
=
e^{-2\pi i\alpha y}
e^{2\pi i k y/N_y}
\ket{0,y}.
\label{eq:app-Tx-boundary}
\end{equation}

\subsection*{Magnetic translation algebra}

In the bulk,
\begin{align}
T_xT_y\ket{x,y}
&=
T_x\ket{x,y+1}
\nonumber\\
&=
e^{-2\pi i\alpha (y+1)}
\ket{x+1,y+1},
\label{eq:app-TxTy}
\end{align}
\begin{align}   
T_yT_x\ket{x,y}
&=
T_y\left(
e^{-2\pi i\alpha y}\ket{x+1,y}
\right)
\nonumber\\
&=
e^{-2\pi i\alpha y}
\ket{x+1,y+1}.
\label{eq:app-TyTx}
\end{align}
Comparing the two expressions gives
\begin{equation}
T_xT_y
=
e^{-2\pi i\alpha}
T_yT_x .
\label{eq:app-magnetic-algebra}
\end{equation}
Equivalently,
\begin{equation}
T_yT_x
=
e^{2\pi i\alpha}
T_xT_y .
\end{equation}
The sign in the phase depends only on the convention chosen for $T_x$; the
projective content is the same.

\subsection*{Bulk commutator with $T_x$}

We first check the commutator with $T_x$ away from the boundary. Starting
from \eqref{eq:app-h-bulk}, we have
\begin{align}
T_xh\ket{x,y}
=
-t\Big[
&T_x\ket{x-1,y}
+T_x\ket{x+1,y}
\nonumber\\
&+e^{2\pi i\alpha x}T_x\ket{x,y-1}
+e^{-2\pi i\alpha x}T_x\ket{x,y+1}
\Big].
\end{align}
Using \eqref{eq:app-Tx-bulk},
\begin{align}
T_xh\ket{x,y}
=
-t\Big[
&e^{-2\pi i\alpha y}\ket{x,y}
+e^{-2\pi i\alpha y}\ket{x+2,y}
\nonumber\\
&+e^{2\pi i\alpha x}
 e^{-2\pi i\alpha (y-1)}
 \ket{x+1,y-1}
\nonumber\\
&+e^{-2\pi i\alpha x}
 e^{-2\pi i\alpha (y+1)}
 \ket{x+1,y+1}
\Big].
\label{eq:app-Txh-bulk}
\end{align}
On the other hand,
\begin{align}
hT_x\ket{x,y}
&=
h\left(e^{-2\pi i\alpha y}\ket{x+1,y}\right)
\nonumber\\
&=
e^{-2\pi i\alpha y}h\ket{x+1,y}
\nonumber\\
&=
-t\Big[
e^{-2\pi i\alpha y}\ket{x,y}
+e^{-2\pi i\alpha y}\ket{x+2,y}
\nonumber\\
&\hspace{1.8cm}
+e^{-2\pi i\alpha y}
 e^{2\pi i\alpha (x+1)}
 \ket{x+1,y-1}
\nonumber\\
&\hspace{1.8cm}
+e^{-2\pi i\alpha y}
 e^{-2\pi i\alpha (x+1)}
 \ket{x+1,y+1}
\Big].
\label{eq:app-hTx-bulk}
\end{align}
The third coefficient in \eqref{eq:app-Txh-bulk} is
\begin{equation}
e^{2\pi i\alpha x}e^{-2\pi i\alpha(y-1)}
=
e^{2\pi i\alpha x}e^{-2\pi i\alpha y}e^{2\pi i\alpha}
=
e^{-2\pi i\alpha y}e^{2\pi i\alpha(x+1)},
\end{equation}
which equals the third coefficient in \eqref{eq:app-hTx-bulk}. Similarly,
\begin{equation}
e^{-2\pi i\alpha x}e^{-2\pi i\alpha(y+1)}
=
e^{-2\pi i\alpha y}e^{-2\pi i\alpha(x+1)}.
\end{equation}
Therefore
\begin{equation}
T_xh\ket{x,y}=hT_x\ket{x,y}
\end{equation}
in the bulk.

\subsection*{Boundary commutator with $T_x$}

We now check the boundary state $\ket{N_x-1,y}$.
First, we compute $T_xh\ket{N_x-1,y}$. Using
\eqref{eq:app-h-boundary-expanded},
\begin{align}
T_xh\ket{N_x-1,y}
=
-t\Big[
&T_x\ket{N_x-2,y}
+e^{2\pi i k y/N_y}T_x\ket{0,y}
\nonumber\\
&+e^{2\pi i k/N_y-2\pi i\alpha}
 T_x\ket{N_x-1,y-1}
\nonumber\\
&+e^{-2\pi i k/N_y+2\pi i\alpha}
 T_x\ket{N_x-1,y+1}
\Big].
\end{align}
Using the definitions of $T_x$, including the boundary action, gives
\begin{align}
T_xh\ket{N_x-1,y}
=
-t\Big[
&e^{-2\pi i\alpha y}\ket{N_x-1,y}
\nonumber\\
&+e^{2\pi i k y/N_y}
 e^{-2\pi i\alpha y}
 \ket{1,y}
\nonumber\\
&+e^{2\pi i k/N_y-2\pi i\alpha}
 e^{-2\pi i\alpha (y-1)}
 e^{2\pi i k (y-1)/N_y}
 \ket{0,y-1}
\nonumber\\
&+e^{-2\pi i k/N_y+2\pi i\alpha}
 e^{-2\pi i\alpha (y+1)}
 e^{2\pi i k (y+1)/N_y}
 \ket{0,y+1}
\Big].
\label{eq:app-Txh-boundary-raw}
\end{align}
We simplify each coefficient. The coefficient of $\ket{1,y}$ is
\begin{equation}
e^{2\pi i k y/N_y}e^{-2\pi i\alpha y}.
\end{equation}
The coefficient of $\ket{0,y-1}$ is
\begin{align}
&e^{2\pi i k/N_y-2\pi i\alpha}
 e^{-2\pi i\alpha (y-1)}
 e^{2\pi i k (y-1)/N_y}
\nonumber\\
&=
e^{-2\pi i\alpha y}
e^{2\pi i k y/N_y}.
\end{align}
The coefficient of $\ket{0,y+1}$ is
\begin{align}
&e^{-2\pi i k/N_y+2\pi i\alpha}
 e^{-2\pi i\alpha (y+1)}
 e^{2\pi i k (y+1)/N_y}
\nonumber\\
&=
e^{-2\pi i\alpha y}
e^{2\pi i k y/N_y}.
\end{align}
Hence
\begin{align}
T_xh\ket{N_x-1,y}
=
-t\Big[
&e^{-2\pi i\alpha y}\ket{N_x-1,y}
\nonumber\\
&+e^{-2\pi i\alpha y}e^{2\pi i k y/N_y}\ket{1,y}
\nonumber\\
&+e^{-2\pi i\alpha y}e^{2\pi i k y/N_y}\ket{0,y-1}
\nonumber\\
&+e^{-2\pi i\alpha y}e^{2\pi i k y/N_y}\ket{0,y+1}
\Big].
\label{eq:app-Txh-boundary}
\end{align}

Now we turn to $hT_x\ket{N_x-1,y}$. From \eqref{eq:app-Tx-boundary},
\begin{align}
hT_x\ket{N_x-1,y}
&=
h\left(
e^{-2\pi i\alpha y}e^{2\pi i k y/N_y}\ket{0,y}
\right)
\nonumber\\
&=
e^{-2\pi i\alpha y}e^{2\pi i k y/N_y}
h\ket{0,y}.
\end{align}
At $x=0$, the backward link goes through the patch in the inverse direction.
Therefore,
\begin{align}
h\ket{0,y}
=
-t\Big[
&e^{-2\pi i k y/N_y}\ket{N_x-1,y}
+\ket{1,y}
\nonumber\\
&+\ket{0,y-1}
+\ket{0,y+1}
\Big].
\label{eq:app-h-zero}
\end{align}
Thus
\begin{align}
hT_x\ket{N_x-1,y}
=
-t\Big[
&e^{-2\pi i\alpha y}\ket{N_x-1,y}
\nonumber\\
&+e^{-2\pi i\alpha y}e^{2\pi i k y/N_y}\ket{1,y}
\nonumber\\
&+e^{-2\pi i\alpha y}e^{2\pi i k y/N_y}\ket{0,y-1}
\nonumber\\
&+e^{-2\pi i\alpha y}e^{2\pi i k y/N_y}\ket{0,y+1}
\Big].
\label{eq:app-hTx-boundary}
\end{align}
Comparing \eqref{eq:app-Txh-boundary} and
\eqref{eq:app-hTx-boundary}, we obtain
\begin{equation}
T_xh\ket{N_x-1,y}
=
hT_x\ket{N_x-1,y}.
\end{equation}
Combining the bulk and boundary results,
\begin{equation}
[h,T_x]\ket{x,y}=0
\qquad
\forall\,\ket{x,y}.
\label{eq:app-Tx-symmetry}
\end{equation}

\subsection*{Bulk commutator with $T_y$}

We now check $T_y$. In the bulk,
\begin{align}
T_yh\ket{x,y}
=
-t\Big[
&\ket{x-1,y+1}
+\ket{x+1,y+1}
\nonumber\\
&+e^{2\pi i\alpha x}\ket{x,y}
+e^{-2\pi i\alpha x}\ket{x,y+2}
\Big].
\label{eq:app-Tyh-bulk}
\end{align}
On the other hand,
\begin{align}
hT_y\ket{x,y}
&=
h\ket{x,y+1}
\nonumber\\
&=
-t\Big[
\ket{x-1,y+1}
+\ket{x+1,y+1}
\nonumber\\
&\hspace{1.8cm}
+e^{2\pi i\alpha x}\ket{x,y}
+e^{-2\pi i\alpha x}\ket{x,y+2}
\Big].
\label{eq:app-hTy-bulk}
\end{align}
Therefore
\begin{equation}
T_yh\ket{x,y}=hT_y\ket{x,y}
\end{equation}
in the bulk.

\subsection*{Boundary commutator with $T_y$}

At the boundary $x=N_x-1$, using \eqref{eq:app-h-boundary-expanded},
\begin{align}
T_yh\ket{N_x-1,y}
=
-t\Big[
&\ket{N_x-2,y+1}
+e^{2\pi i k y/N_y}\ket{0,y+1}
\nonumber\\
&+e^{2\pi i k/N_y-2\pi i\alpha}\ket{N_x-1,y}
\nonumber\\
&+e^{-2\pi i k/N_y+2\pi i\alpha}\ket{N_x-1,y+2}
\Big].
\label{eq:app-Tyh-boundary}
\end{align}
Instead,
\begin{align}
hT_y\ket{N_x-1,y}
&=
h\ket{N_x-1,y+1}
\nonumber\\
=
-t\Big[
&\ket{N_x-2,y+1}
+e^{2\pi i k (y+1)/N_y}\ket{0,y+1}
\nonumber\\
&+e^{2\pi i k/N_y-2\pi i\alpha}\ket{N_x-1,y}
\nonumber\\
&+e^{-2\pi i k/N_y+2\pi i\alpha}\ket{N_x-1,y+2}
\Big].
\label{eq:app-hTy-boundary}
\end{align}
The two expressions differ only in the coefficient of $\ket{0,y+1}$:
\begin{equation}
e^{2\pi i k y/N_y}
\neq
e^{2\pi i k (y+1)/N_y}
\end{equation}
unless $e^{2\pi i k/N_y}=1$. Therefore the elementary translation $T_y$
is not, in general, an exact symmetry of the finite magnetic torus:
\begin{equation}
[h,T_y]\ket{N_x-1,y}\neq0.
\label{eq:app-Ty-not-symmetry}
\end{equation}
More explicitly,
\begin{align}
\left(hT_y-T_yh\right)\ket{N_x-1,y}
=
-t\left[
e^{2\pi i k (y+1)/N_y}
-
e^{2\pi i k y/N_y}
\right]\ket{0,y+1}.
\end{align}
Thus the obstruction is entirely due to the magnetic transition function at
the $x$-boundary.

\subsection*{Reduced magnetic translations}

We now impose the commensurability conditions
\begin{equation}
k\mid N_x,
\qquad
k\mid N_y.
\end{equation}
We define
\begin{equation}
m_x=\frac{N_x}{k},
\qquad
m_y=\frac{N_y}{k},
\end{equation}
and introduce the reduced magnetic translations
\begin{equation}
X=(T_x)^{m_x},
\qquad
Z=(T_y)^{m_y}.
\label{eq:app-XZ}
\end{equation}

Since $T_x$ commutes with $h$ everywhere, it follows immediately that
\begin{equation}
[h,X]=0.
\label{eq:app-X-commutes}
\end{equation}
The non-trivial point is the commutator of $Z$ with $h$ at the boundary.

\subsection*{Boundary check for $Z=(T_y)^{m_y}$}

We compute $Zh\ket{N_x-1,y}$ and $hZ\ket{N_x-1,y}$.
First,
\begin{align}
Zh\ket{N_x-1,y}
=
(T_y)^{m_y}h\ket{N_x-1,y}.
\end{align}
Using \eqref{eq:app-h-boundary-expanded},
\begin{align}
Zh\ket{N_x-1,y}
=
-t\Big[
&\ket{N_x-2,y+m_y}
+e^{2\pi i k y/N_y}\ket{0,y+m_y}
\nonumber\\
&+e^{2\pi i k/N_y-2\pi i\alpha}
 \ket{N_x-1,y+m_y-1}
\nonumber\\
&+e^{-2\pi i k/N_y+2\pi i\alpha}
 \ket{N_x-1,y+m_y+1}
\Big].
\label{eq:app-Zh-boundary}
\end{align}
On the other hand,
\begin{align}
hZ\ket{N_x-1,y}
&=
h\ket{N_x-1,y+m_y}
\nonumber\\
=
-t\Big[
&\ket{N_x-2,y+m_y}
+e^{2\pi i k (y+m_y)/N_y}\ket{0,y+m_y}
\nonumber\\
&+e^{2\pi i k/N_y-2\pi i\alpha}
 \ket{N_x-1,y+m_y-1}
\nonumber\\
&+e^{-2\pi i k/N_y+2\pi i\alpha}
 \ket{N_x-1,y+m_y+1}
\Big].
\label{eq:app-hZ-boundary}
\end{align}
The only possible mismatch is again the boundary coefficient:
\begin{equation}
e^{2\pi i k (y+m_y)/N_y}
=
e^{2\pi i k y/N_y}
e^{2\pi i k m_y/N_y}.
\end{equation}
But since
\begin{equation}
m_y=\frac{N_y}{k},
\end{equation}
we have
\begin{equation}
e^{2\pi i k m_y/N_y}
=
e^{2\pi i}=1.
\end{equation}
Therefore
\begin{equation}
e^{2\pi i k (y+m_y)/N_y}
=
e^{2\pi i k y/N_y},
\end{equation}
and hence
\begin{equation}
hZ\ket{N_x-1,y}
=
Zh\ket{N_x-1,y}.
\end{equation}
Combining this with the bulk commutation, we obtain
\begin{equation}
[h,Z]\ket{x,y}=0
\qquad
\forall\,\ket{x,y}.
\label{eq:app-Z-commutes}
\end{equation}

Thus the reduced translations satisfy
\begin{equation}
[h,X]=[h,Z]=0.
\label{eq:app-reduced-symmetries}
\end{equation}

\subsection*{Projective algebra of the reduced translations}

Using the elementary magnetic translation algebra
\eqref{eq:app-magnetic-algebra},
\begin{equation}
T_xT_y=e^{-2\pi i\alpha}T_yT_x,
\end{equation}
we compute
\begin{align}
XZ
&=
T_x^{m_x}T_y^{m_y}.
\end{align}
Each time a $T_x$ is moved past a $T_y$, one obtains a phase
$e^{-2\pi i\alpha}$. Since there are $m_xm_y$ such interchanges,
\begin{align}
T_x^{m_x}T_y^{m_y}
&=
e^{-2\pi i\alpha m_xm_y}
T_y^{m_y}T_x^{m_x}.
\end{align}
Therefore
\begin{equation}
XZ
=
e^{-2\pi i\alpha m_xm_y}
ZX.
\end{equation}
Using
\begin{align}
\alpha m_xm_y
&=
\frac{k}{N_xN_y}
\frac{N_x}{k}
\frac{N_y}{k}
\nonumber\\
&=
\frac{1}{k},
\end{align}
we obtain
\begin{equation}
XZ=e^{-2\pi i/k}ZX.
\label{eq:app-projective-algebra}
\end{equation}
Equivalently, depending on the convention for $T_x$, one may write
$XZ=e^{2\pi i/k}ZX$. The sign is conventional; the important point is that
$X$ and $Z$ generate the finite Heisenberg algebra at level $k$.

\subsection*{Origin of the exact degeneracy}

The reduced translations $X$ and $Z$ are exact symmetries of the
Hamiltonian and obey
\begin{equation}
XZ=e^{-2\pi i/k}ZX.
\end{equation}
This is the defining projective algebra of the finite magnetic translation
group. Its irreducible representations have dimension $k$.

Indeed, if $\ket{\psi}$ is an eigenstate of $h$ and also an eigenstate of
$Z$,
\begin{equation}
h\ket{\psi}=E\ket{\psi},
\qquad
Z\ket{\psi}=z\ket{\psi},
\end{equation}
then
\begin{equation}
h(X\ket{\psi})
=
Xh\ket{\psi}
=
E(X\ket{\psi}),
\end{equation}
so $X\ket{\psi}$ has the same energy. Moreover,
\begin{align}
Z(X\ket{\psi})
&=
ZX\ket{\psi}
\nonumber\\
&=
e^{2\pi i/k}XZ\ket{\psi}
\nonumber\\
&=
e^{2\pi i/k}z\,X\ket{\psi}.
\end{align}
Thus repeated action of $X$ generates states with the same energy but with
distinct $Z$-eigenvalues
\begin{equation}
z,
\quad e^{2\pi i/k}z,
\quad e^{4\pi i/k}z,
\quad \ldots,
\quad e^{2\pi i(k-1)/k}z.
\end{equation}
After $k$ applications one returns to the original eigenvalue. Therefore the
multiplet has dimension $k$.

This proves that, whenever
\begin{equation}
k\mid N_x,
\qquad
k\mid N_y,
\end{equation}
the spectrum displays an exact $k$-fold degeneracy generated by the reduced
magnetic translations $X$ and $Z$.

\acknowledgments

The authors gratefully acknowledge Stefan K\"{u}hn for insightful comments and collaboration at different stages of this work. M.C.D. thanks CERN's hospitality while this work was done. A.B., E.R.O. and S.S. acknowledge Matteo Wauters forsuggesting the study of the Hofstadter butterfly in our setup and Alessio Celi for useful discussions. The authors gratefully acknowledge the granted access to the Marvin cluster hosted by the University of Bonn. This project was supported by the Deutsche Forschungsgemeinschaft (DFG, German Research Foundation) as part of the CRC 1639 NuMeriQS -- project no.\ 511713970 and under Germany's Excellence Strategy -- Cluster of Excellence ``Color meets Flavor'' (CmF) EXC 3107 -- 533766364 and Cluster of Excellence ``Matter and Light for Quantum Computing'' (ML4Q) EXC 2004/2 -- 390534769. This work is part of the Quantum Computing for High-Energy Physics (QC4HEP) working group. This work has been partially funded by the Eric \& Wendy Schmidt Fund for Strategic Innovation through the CERN Next Generation Triggers project under grant agreement number SIF-2023-004. This work is supported with funds from the Ministry of Science, Research and Culture of the State of Brandenburg within the Center for Quantum Technologies and Applications (CQTA). 
\begin{center}

    \includegraphics[width = 0.1\textwidth]{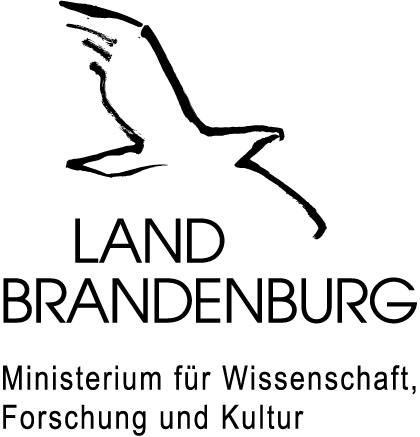}

\end{center}
This work is funded by the European Union’s Horizon Europe Framework Programme (HORIZON) under the ERA Chair scheme with grant agreement no. 101087126.

\bibliographystyle{JHEP}
\bibliography{references}

\end{document}